\documentclass[a4paper,fleqn,usenatbib]{mnras}

\usepackage{newtxtext,newtxmath}
\usepackage{natbib}

\usepackage[T1]{fontenc}
\usepackage{ae,aecompl}

\usepackage{graphicx}	
\usepackage{amsmath}	

\usepackage{amssymb}	
\usepackage{supertabular}
\usepackage{cuted}
\usepackage{array}
\usepackage{subfigure}

\title[State Transitions of GX 339-4]{State transitions of GX 339-4 during its outburst rising phase}

\author[Q. C. Shui et al.]
{Q. C. Shui,$^{1,2,3}$
H. X. Yin,$^{1}$\thanks{E-mail: yinhx@sdu.edu.cn}
S. Zhang,$^{2}$
J. L. Qu,$^{2,3}$
Y. P. Chen,$^{2}$
L. D. Kong,$^{2,3}$
P. J. Wang,$^{2,3}$
\newauthor
H. F. Zhang,$^{1}$
J. X. Song,$^{1}$
B. Ning,$^{1}$
Y. F. Wang,$^{1}$
Z. Chang$^{2}$
and
P. Zhang$^{2}$
\\
$^{1}$Shandong Key Laboratory of Optical Astronomy and Solar-Terrestrial Environment, School of Space Science and Physics, Institute of Space Sciences, \\  Shandong University, Weihai, Shandong, 264209, China\\
$^{2}$Key Laboratory of Particle Astrophysics, Institute of High Energy Physics, Chinese Academy of Sciences, 100049, Beijing, China\\
$^{3}$University of Chinese Academy of Sciences, Chinese Academy of Sciences, 100049, Beijing, China\\
}

\date{Accepted 2021 September 1. Received 2021 July 31; in original form 2020 October 27}

\pubyear{2021}

\begin{document}
\label{firstpage}
\pagerange{\pageref{firstpage}--\pageref{lastpage}}
\maketitle

\begin{abstract}
We investigate systematically four outbursts of black hole system GX 339-4 observed by the \textit{Rossi X-ray Timing Explorer} (\textit{RXTE}) in both spectral and timing domains and find that these outbursts have some common properties although they experience different ‘q’ tracks in the hardness-intensity diagram (HID).  
While the spectral indices are around 1.5 in low hard state (LHS), 2.4 in soft intermediate state (SIMS) and high soft state (HSS), the spectral parameters of thermal, non-thermal and reflection components vary significantly in transitions from LHS to HIMS.  
Also the quasi periodic oscillation (QPO) shows a peculiar behavior during the state transition between LHS and HIMS: the RMS drop of type C fundamental QPO is accompanied with showing-up of the second harmonic.
Interestingly, the QPO RMS is found to have a similar linear relationship with the non-thermal fraction of emission in different outbursts.  
These findings provide more clues to our understanding the outburst of the black hole X-ray binary system.
\end{abstract}




\section{INTRODUCTION}
Black hole binaries (BHBs) undergo outbursts occasionally via accretion from the companion. 
The X-ray spectrum of BHBs during outburst usually consists of thermal and non-thermal components: while the thermal is believed to be the disk emission \citep{1973A&A....24..337S,1973blho.conf..343N,1974MNRAS.168..603L}, the non-thermal may have different origins, which can be Comptonizations in either the corona or jet, both with seed photons fed by the disk \citep{1980A&A....86..121S,1985A&A...143..374S,1994ApJ...434..570T,1995ApJ...452..710N,1997ApJ...489..865E,2014ARA&A..52..529Y}. Part of the non-thermal emissions interact with the disk and end up as the reflection component (\citealp{2014ApJ...782...76G,2015ApJ...808L..37G} and references therein).
The BHB outbursts are observed to have four canonical states traced in hardness-intensity diagram (HID, \citealp{1990A&A...235..131H,2005A&A...440..207B,2005Ap&SS.300..107H,2006AdSpR..38.2801B,2006ARA&A..44...49R,2009MNRAS.396.1370F,2011BASI...39..409B}): low/hard state (LHS), hard intermediate state (HIMS), soft intermediate (SIMS) and high/soft state (HSS). The outburst evolution is characterized by strong variabilities of the non-thermal emission in LHS, softened spectrum and inward movement of disk in HIMS \citep{1997ApJ...489..865E,2014ARA&A..52..529Y}, even softer spectrum in SIMS, and thermal domination in HSS.

In the timing domain, the most remarkable features of the outburst in power density spectrum (PDS) are quasi-periodic oscillations (QPOs, \citealp{1989ARA&A..27..517V}). Low-frequency QPOs (LFQPOs; roughly 0.1 $\sim$ 30 Hz) are classified into type A, B, and C on the basis of their centroid frequencies, time lags, and total RMS amplitudes \citep{1999ApJ...526L..33W,2001ApJS..132..377H,2006ARA&A..44...49R}. QPOs are usually observed in the HSS for type A, characterized by a weak and broad peak at frequency around 6 $\sim$ 8 Hz, in the SIMS for type B, characterized by a relatively strong fundamental and a weak harmonic, and in LHS/HIMS for type C, characterized by strong flat-top noise component in PDS \citep{2011BASI...39..409B,2016ASSL..440...61B}. One popular interpretation of LFQPOs is the so-called Lense-Thirring (LT) precession model: the precession of the entire inner accretion flow can produce QPO features and the propagating fluctuations can be responsible for the broad-band noise \citep{2009MNRAS.397L.101I,2011MNRAS.415.2323I}. Further evidences to this model come from energy dependences of the LFQPOs (see \citealt{1997ApJ...482..993M,1999ApJ...524L..59C,2004ApJ...615..416R,2010ApJ...710..836Q,2012Ap&SS.337..137Y,2013MNRAS.433..412L,2018ApJ...866..122H}).

GX 339-4 is a transient X-ray binary \citep{2003ApJ...583L..95H} discovered in 1971 \citep{1973ApJ...184L..67M}. It becomes one of the well studied BHBs due to a series of outbursts observed so far. Based on the absorption lines in near-infrared band, \citet{2017ApJ...846..132H} identified the companion as a giant K-type star.
The inclination angle of the system is still in debate in a range of $30^{\circ} \sim 60^{\circ}$ \citep{2002AJ....123.1741C,2016ApJ...821L...6P,2019MNRAS.488.1026Z}.
The mass function was measured as $f( M) = 5.8\pm0.5 {M_{\sun}}$ \citep{2003ApJ...583L..95H} and the BH mass was estimated as $ 8.28 M_{\sun}$ $\sim$ $ 11.89 M_{\sun}$ \citep{2019AdSpR..63.1374S}. The harbored BH has a spin of $ a = 0.93 \pm 0.01$ \citep{2008ApJ...679L.113M} or $a = 0.935 \pm 0.01$ measured by \textit{XMM-Newton} and \textit{RXTE} \citep{2008MNRAS.387.1489R}.

In history, a series of outbursts of GX 339-4 have been observed and investigated in details. \citet{2009MNRAS.400.1603M} found in 2007 outburst that the cut-off energy decreased in LHS, then increased rapidly in HIMS and SIMS, and finally remained constant in HSS. Also the spectral index increased smoothly in LHS but faster in HIMS. \citet{2015MNRAS.447.1984D} found the thermal emission went up faster after the LHS-HIMS transition during the 2010 outburst of GX 339-4, accompanied with the sudden rise of the normalization of the power-law component in the HIMS \citep{2019MNRAS.486.2705A}. The harmonic of type C QPOs are usually observed in HIMS but not LHS \citep{2016MNRAS.458.1778A,2017ApJ...845..143Z}, and type B QPOs usually occur at the end of HIMS \citep{2004A&A...426..587C,2016ASSL..440...61B}. 
The radio jet during outbursts is speculated to be sporadic in SIMS but stable in LHS and HIMS \citep{2006MNRAS.367.1432M}. These show the complexity of outburst, which needs systematical investigations on a series of outburst samples.

In this paper, we investigate in details four outbursts of GX 339-4, focusing on the state transitions in the rising phase. We introduce observations and data reductions in Section~\ref{sec:2}, present results in Section~\ref{sec:3}, discuss in Section~\ref{sec:4} and finally summarize in Section~\ref{sec:5}.

\section{OBSERVATIONS AND DATA REDUCTION}
\label{sec:2}
\begin{figure}
	\includegraphics[width=\columnwidth]{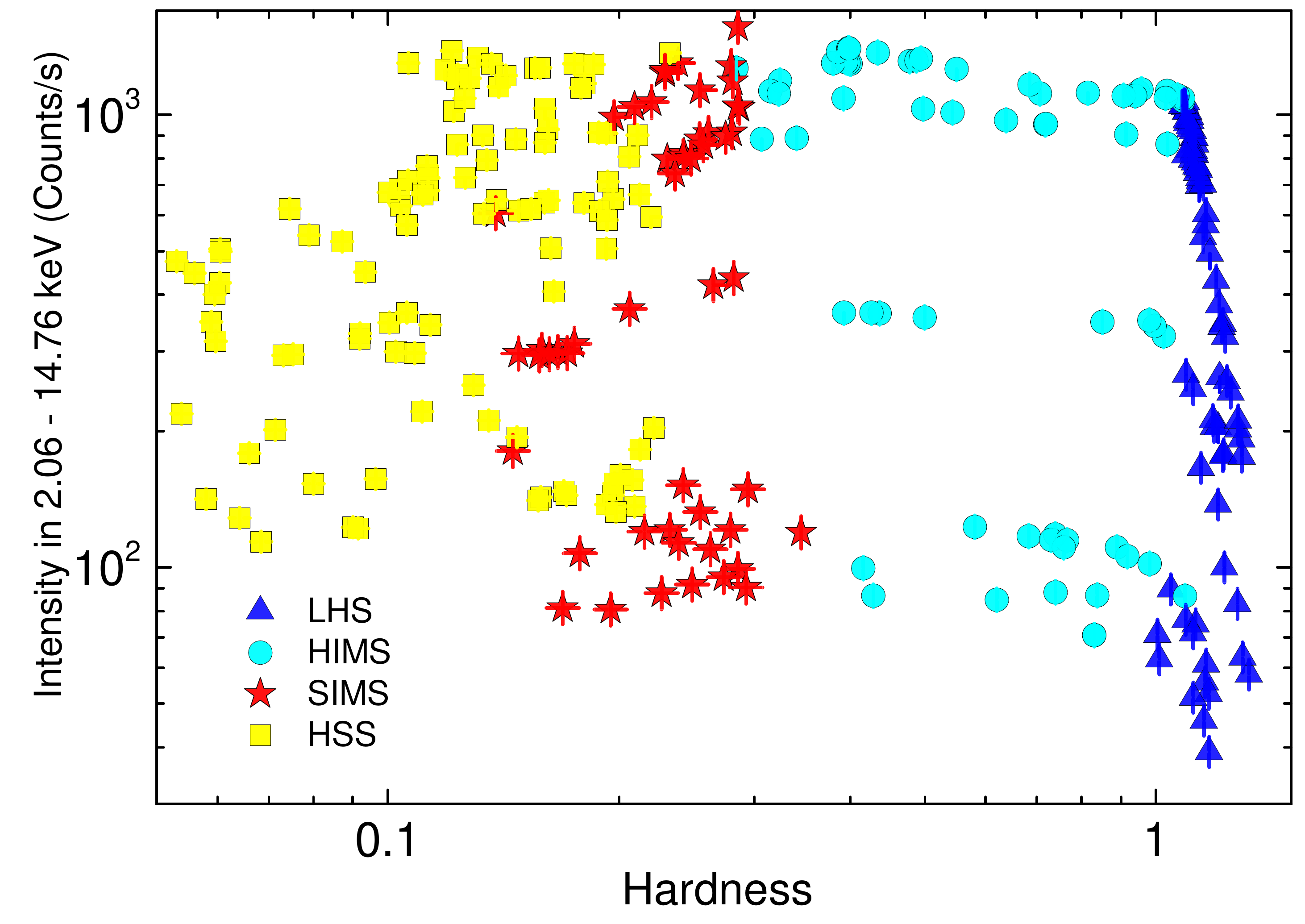}
	\caption{Hardness-intensity diagrams of all observations in our study. Each point represents a single RXTE observation. Here, different symbols and colors represent different spectral states (blue triangles: LHS, cyan circles: HIMS, red stars: SIMS and yellow squares: HSS, respectively). The spectral states are defined by us following \citet{2011MNRAS.418.2292M}.}
	\label{fig:HID}
\end{figure}

We have systematically analyzed observations of GX 339-4 carried out by \textit{RXTE} satellite during the 2002, 2004, 2007 and 2010 outbursts, mainly focusing on the state transitions during the rising phases. The \verb'HEASOFT' software package version 6.24 is adopted and the good time intervals (GTIs) are generated with elevation angle larger than $10^\circ$ and offset less than $0.02^\circ$. The standard data products of e.g. light curve, spectrum and background are produced from data born out of the PCU2 of the \textit{RXTE} Proportional Counter Array (PCA). Hardness is defined as ratio of the count rate at energies between 5.71 $\sim$ 14.76 keV and 2.06 $\sim$ 5.71 keV. And the intensity in the HID takes the PCU2 count rate in the energy range 2.06 $\sim$ 14.76 keV. The HID of four outbursts are presented in Figure~\ref{fig:HID}. The spectral states are distinguished following \citet{2011MNRAS.418.2292M}. Standard 2 mode of PCA data are adopted for spectral analysis in the energy range 2.5$\sim$30.0 keV without any groupings or binnings. To
account for calibration uncertainties, a systematic error of 0.5\% is added in spectral fittings. The spectral analysis is performed using the XSPEC version 12.9.1. The 90\%-confident-level uncertainties are
computed by Markov Chain Monte Carlo (MCMC) with a length 20000.

\begin{figure*}
	\includegraphics[width=18cm]{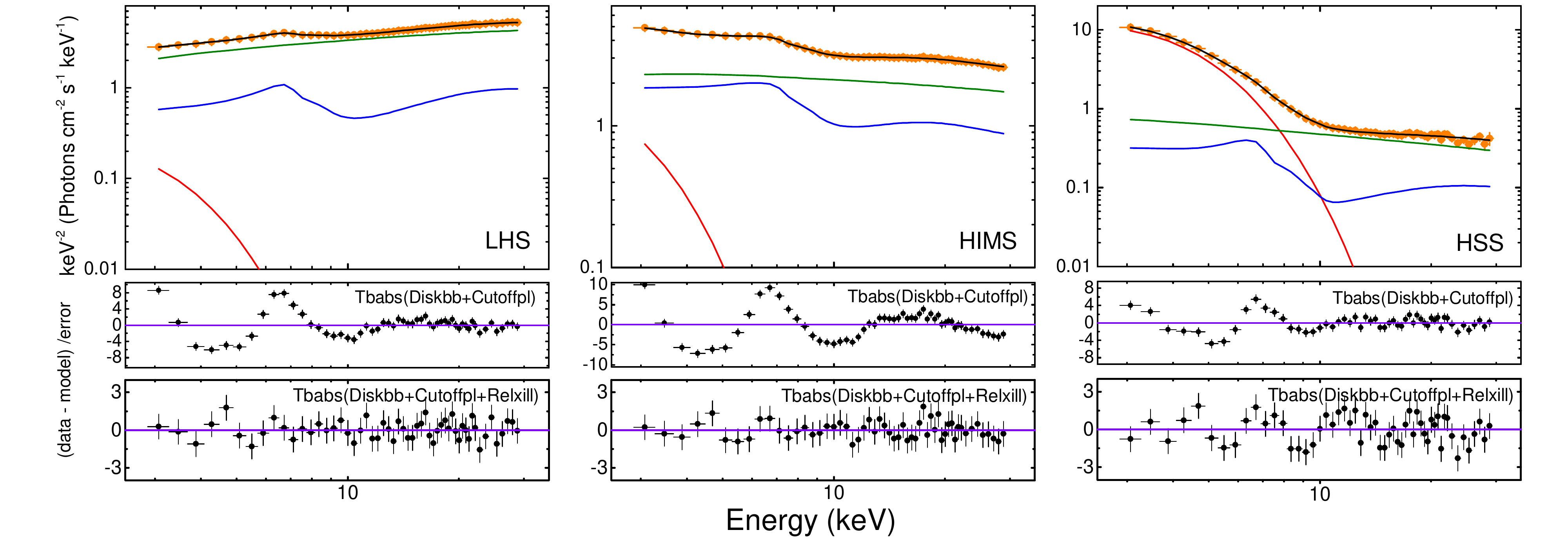}
	\caption{Unfolded spectrum (top) and residuals (middle and bottom) of three states (from left to right followed as: LHS, HIMS and HSS).The total, DISKBB, CUTOFFPL and RELXILL model are displayed as black, red, green and blue, respectively.}
	\label{fig_spectrum}
\end{figure*}
	
\section{RESULTS}
\label{sec:3}
\subsection{Energy spectrum}
We try firstly an empirical model consisting of a cutoff power-law, a disk blackbody and an interstellar absorption: (\texttt{TBABS$\times$(DISKBB+CUTOFFPL)}), where $N_{\rm H}$ is set to $\rm 5\times10^{21} cm^{-2}$ following \citet{1997ApJ...479..926M}. 
With such a model, the  reduced $\chi^2$ is in general larger than 2 and, as shown in Figure~\ref{fig_spectrum}, the residuals show obvious structures at energies around iron line and Compton hump. Hence a further component is needed and a reflection model \texttt{RELXILL} \citep{2014ApJ...782...76G} is added additionally. 
We notice that the \texttt{DISKBB} contribution is marginal in LHS, where its parameters of $R_{\rm in}$ and Temperature are not well constrained by using only PCA data, and the spectral fitting doesn't even need this component in some observations (see also \citealt{2014MNRAS.442.1767P}, \citealp{2019MNRAS.486.2705A} and references therein). 
Since the cut-off energy ($E_{\rm cut}$) of the \texttt{CUTOFFPL} influences the reflection hump at around 20$\sim$40 keV, $E_{\rm cut}$ can be estimated with energies beyond the spectral coverage \citep{2015ApJ...808L..37G}. However, for the less constrained $E_{\rm cut}$, we fix it at 300 keV (see also \citealp{2005MNRAS.358..211R} and \citealp{2010ApJ...718..695G}).
To compare the different outbursts the model (\texttt{TBABS$\times$(DISKBB+CUTOFFPL+RELXILL)}) is applied to all observations and the resulted spectral parameters are presented in Table~\ref{tab:3}, where the reduced  $\chi^2$ suggests that the fittings are generally acceptable.

As for the reflection component \texttt{RELXILL}, the reflection fraction parameter is set to -1, so \texttt{RELXILL} only calculate the reflection emission \citep{2020ApJ...890...53S}. For all the other parameters, redshift is set to 0, the photon index ($\Gamma$) and cut-off energy are linked to those in \texttt{CUTOFFPL}, black hole spin is taken as 0.93 and iron abundance adopt as 6.6 in Solar units \citep{2008ApJ...679L.113M,2016ApJ...821L...6P}, the single power-law emissivity index is set to $ r^{-3}$ \citep{1989MNRAS.238..729F} and the inclination angle set to $30^\circ$ \citep{2016ApJ...821L...6P}.
The parameter $ R_{\rm out}$ turns out to be not sensitive to the overall fitting and hence is frozen at the maximum value (1000$R_{\rm g}$, $ R_{\rm g}=GM/c^2$) (see also \citealp{2020ApJ...890...53S}). 

\subsection{Spectral evolution}
\label{sec:3.1}

\begin{figure}
	\includegraphics[width=\columnwidth]{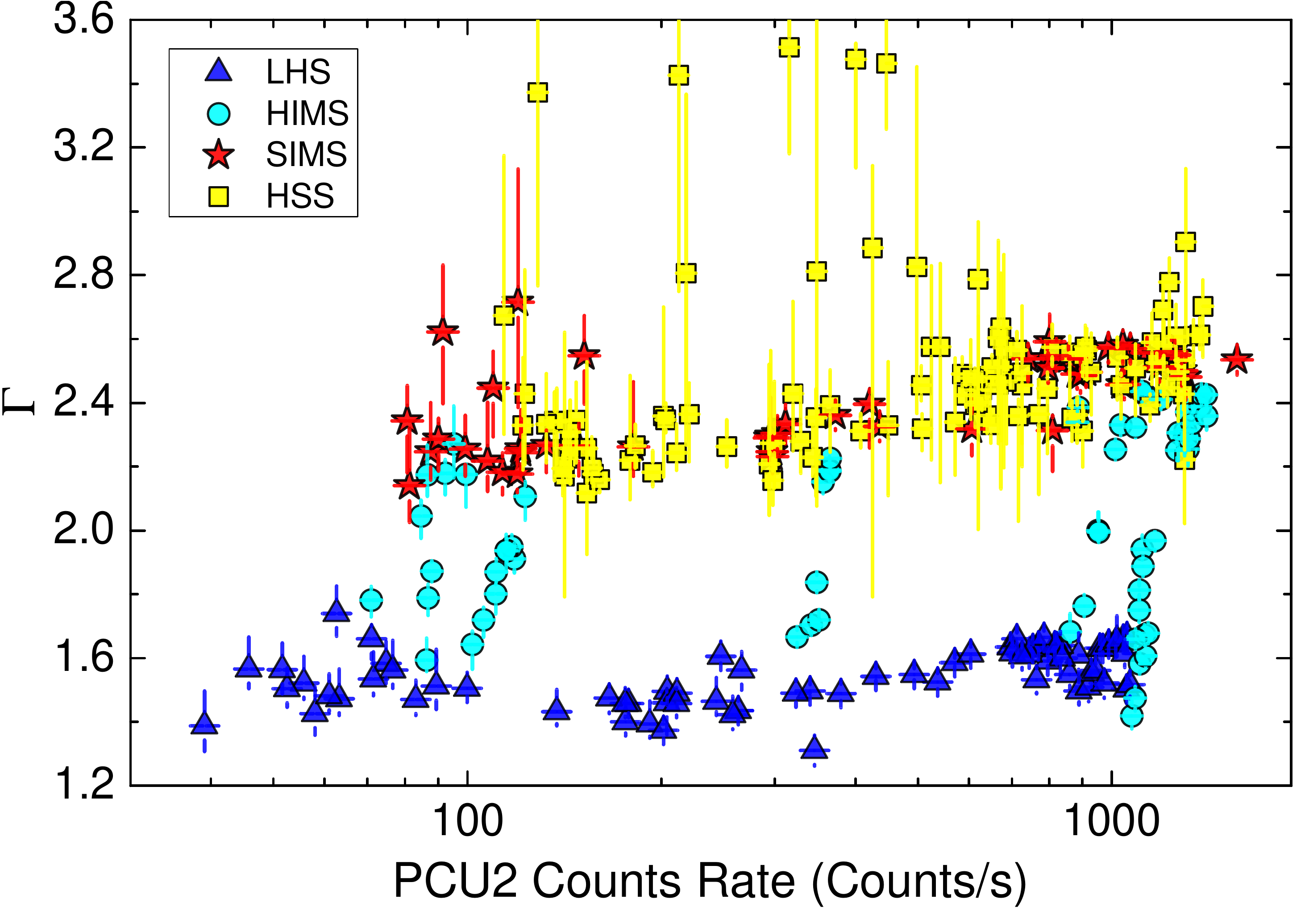}
	\caption{Spectral index versus flux (PCU2 count rate integrated over all energy channels) for four outbursts. Here, different symbols and colors represent different spectral states, having the same meaning as in Figure~\ref{fig:HID}.}
	\label{fig:3}
\end{figure}

\begin{figure*}
	\includegraphics[width=15cm]{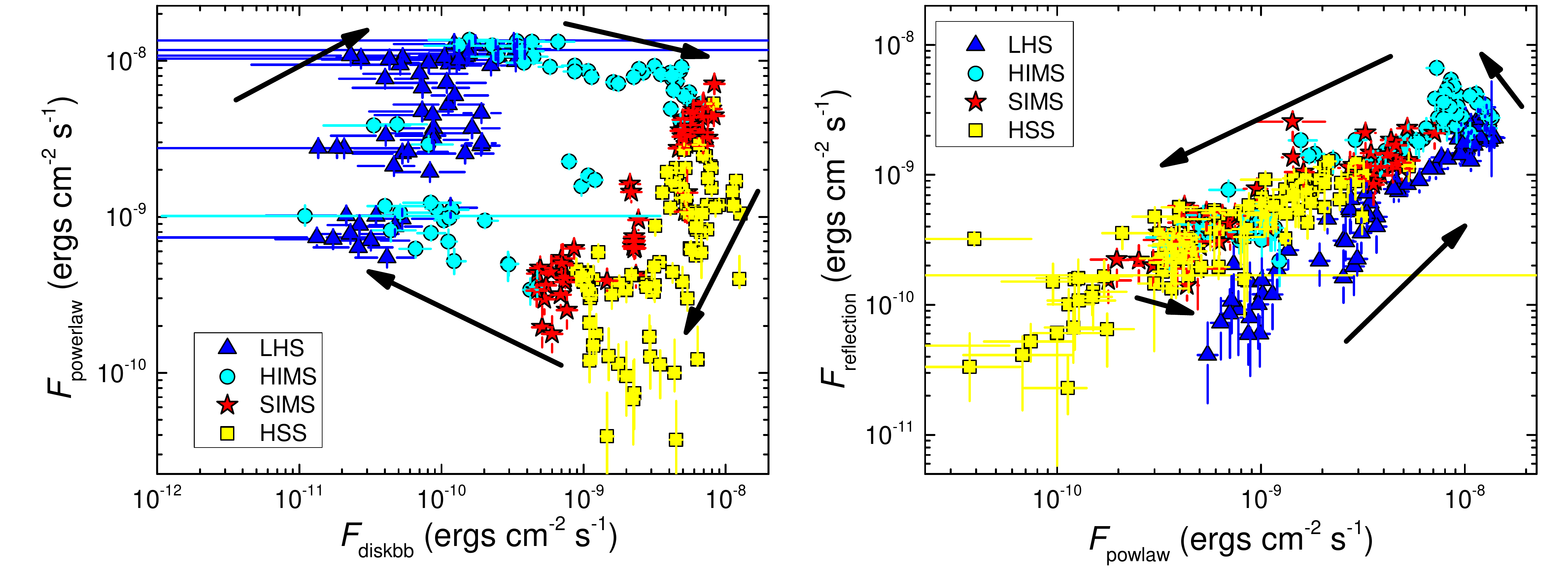}
	\caption{Relations among three component fluxes throughout the four outbursts. Here, different symbols and colors represent different spectral states, having the same meaning as in Figure~\ref{fig:HID} and ~\ref{fig:3}. The directions of arrows indicate the evolution directions of the outbursts.}
	\label{fig:2}
\end{figure*}
\begin{figure}
	\includegraphics[width=\columnwidth]{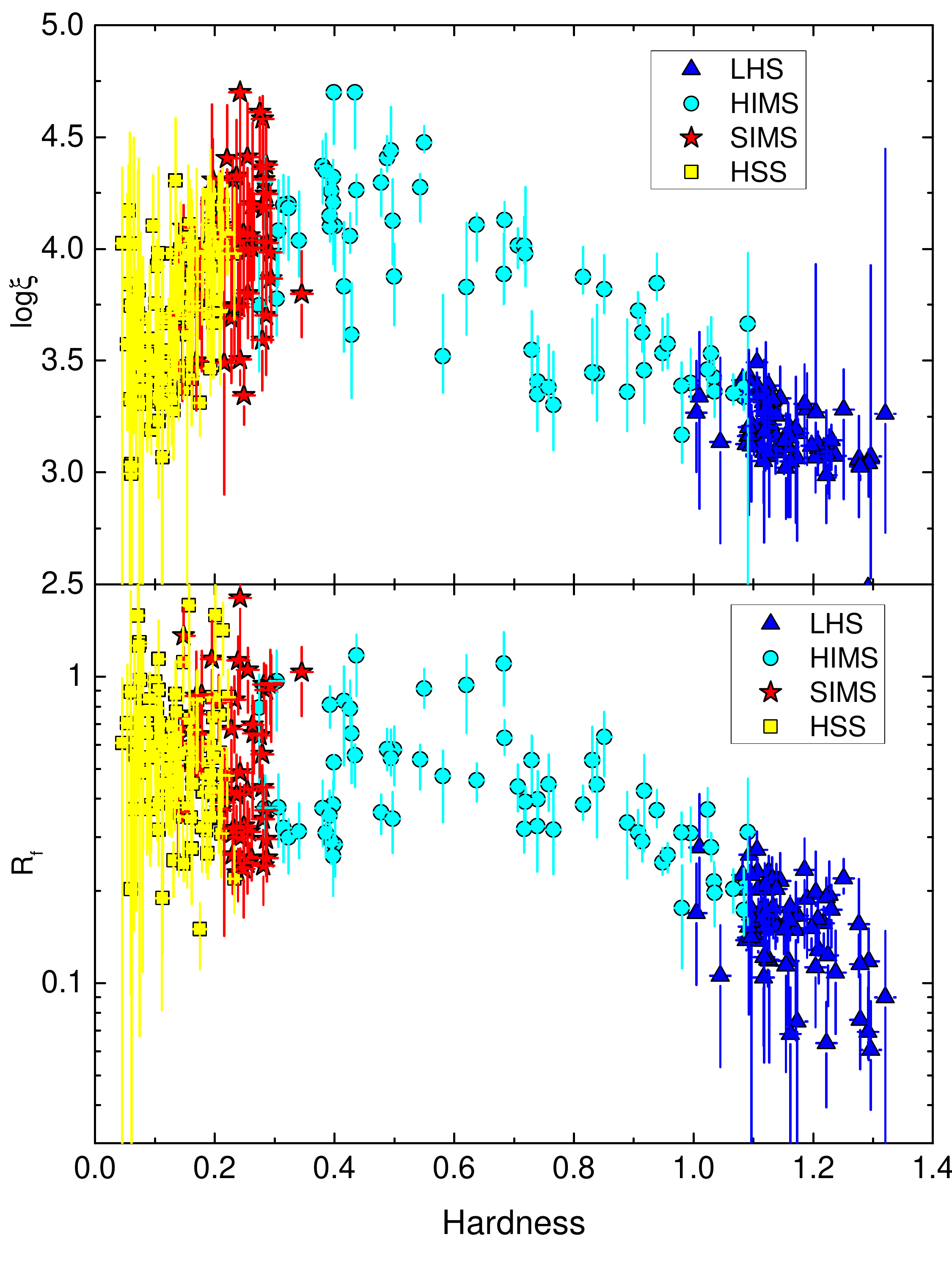}
	\caption{$\log\xi$ and $R_{\rm f}$ plotted against the hardness. Different symbols and colors represent different spectral states, and they have the same meaning as that of Figure~\ref{fig:3}. As one can see, $\log\xi$ and $R_{\rm f}$ increase as the hardness decrease until the source moves into SIMS and HSS.}
	\label{fig:4}
\end{figure}

Although the four outbursts constitute the classic ‘q’ shapes in different regions of the HID diagram (Figure~\ref{fig:HID}), their spectral components and spectral parameters show similarities and diversities. As shown in Figure~\ref{fig:3}, although the state transitions can happen at different flux levels, their spectral indices always start around 1.5 in LHS, and end around 2.4 during SIMS and HSS.
As shown in Figure~\ref{fig:2} (left panel), thermal flux ($ F_{\rm disk}$) is correlated with non-thermal flux ($ F_{\rm pl}$) in LHS, SIMS and HSS, but anti-correlated in HIMS.
The transitions between different spectral states are relevant to the balance between thermal and non-thermal emissions: the stronger outburst is accompanied with higher thermal emissions during transitions of LHS-HIMS and HIMS-SIMS. 
The reflection flux ($ F_{\rm ref}$) increases with $ F_{\rm pl}$ during LHS, SIMS and HSS, while such a trend changes once the source enters HIMS (see the right panel of Figure~\ref{fig:2}). As for the ionization parameter, for all the outbursts it traces the same track in the diagram of $\log\xi$-hardness (the top panel in Figure~\ref{fig:4}). Along with evolution of each outburst, the disk ionization increases all the way until the source moves into SIMS and then decreases in HSS. 
We define the reflection fraction ($R_{\rm f}$) as the ratio of $ F_{\rm ref}$ and $F_{\rm pl}$ and study its evolution with hardness (bottom panel of Figure~\ref{fig:4}). We find $R_{\rm f}$ increases as the source evolves from hard state to soft state and traces the same track in different outbursts.


The evolutions of the spectral components, disk temperature, spectral index, ionization and reflection fraction are plotted for each outburst in Figure~\ref{fig:5}, allowing for investigations of the state transitions LHS-HIMS and HIMS-SIMS in more details. It turns out that, $ F_{\rm pl}$ has the maximum near the LHS-HIMS transition and decreases in the HIMS and SIMS. $ F_{\rm disk}$ is low in LHS and increases fast in HIMS. $F_{\rm ref}$ keeps increasing in the LHS and early HIMS, but drops sharply before the HIMS-SIMS transition. Spectral parameters $\Gamma$, $\log{\xi}$ and $R_{\rm f}$ evolve in a similar manner during LHS and HIMS: keep low in the LHS and then rise fast after the LHS-HIMS transition. Both $T_{\rm in}$ and $\Gamma$ are consistent with being constant in SIMS and HSS, but the ionization and $R_{\rm f}$ start to decrease after the HIMS-SIMS transition.
We note that these behaviors of spectral evolutions are in general consistent with those reported previously \citep[see][]{2009MNRAS.400.1603M,2011MNRAS.418.1746S,2014MNRAS.442.1767P}. Here the spectral analyses provide the essential inputs for joint diagnostic of outbursts with timing analyses, which are shown in what follows.

\subsection{Power spectrum}
The PDS is produced in the 1/32-128 Hz frequency range with 4 ms time resolution by taking Miyamoto normalization \citep{1992ApJ...391L..21M}, from the data of Event and Good xenon modes. The QPOs and other board features in PDS are fitted with Lorentzians. Type C QPOs in the rising phases of 2007 and 2010 outbursts are discriminated following \citet{2011MNRAS.418.2292M}. With such a fitting, QPO parameters are measured for both the fundamental and the second harmonic. The background is corrected with $ RMS = \sqrt{P}\times \left[(S+B)/S\right]$, where $S$ is the source average rate, $B$ the background average rate, and $P$ the power calculated with integration of the QPO Lorentzian function (see \citealp{2015ApJ...799....2B} and \citealp{2020JHEAp..25...29K}). The results derived from PDS fitting are listed in Table~\ref{tab:2}.

\subsection{QPO evolution}

\begin{figure*}
	\includegraphics[width=18cm]{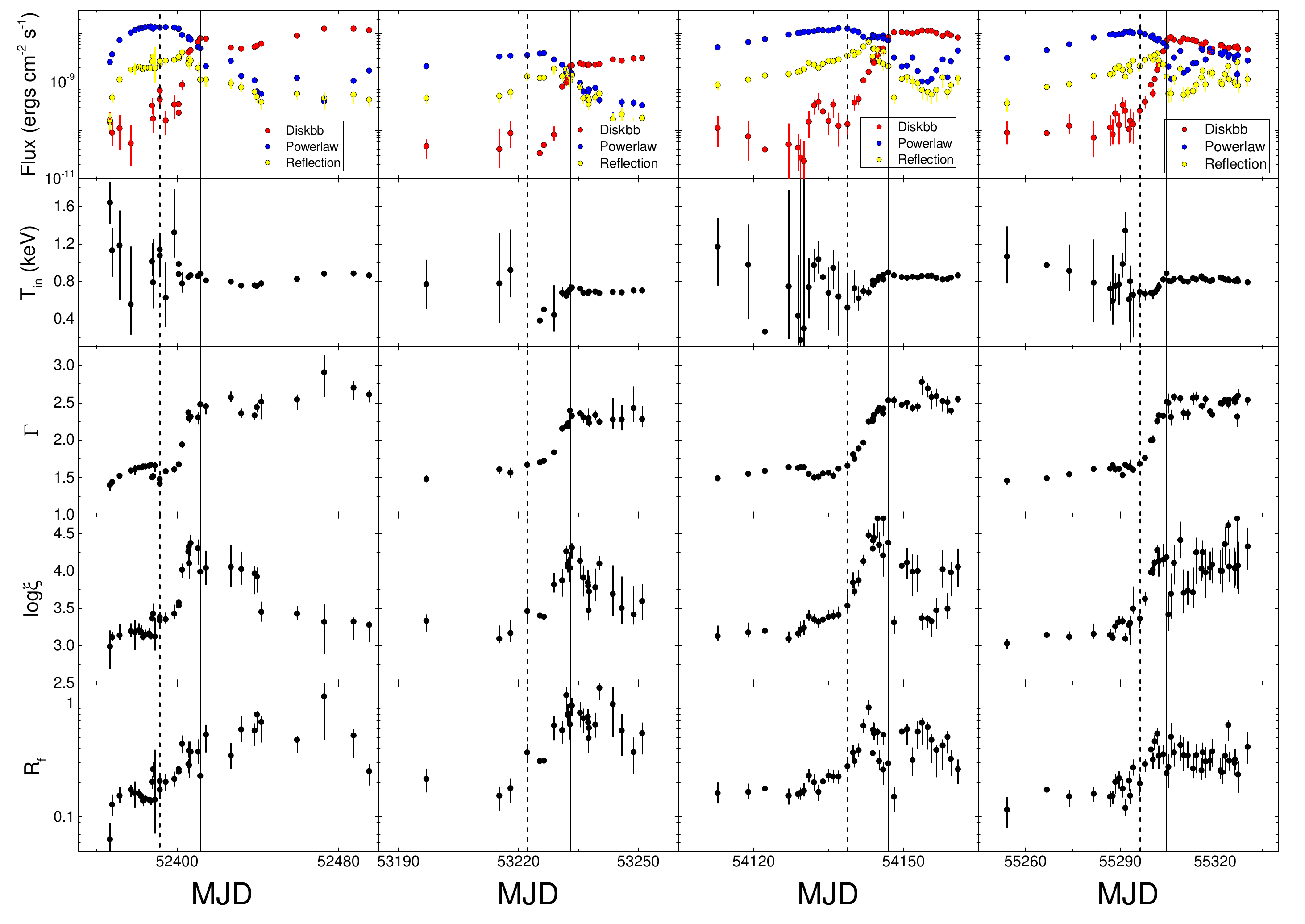}
	\caption{Evolution of the component flux and spectral parameters for observations during the hard-to-soft state transition. From the top panels to the bottom followed as: Logarithm of three model fluxes, disk temperature, photon index, ionization parameter and reflection fraction, respectively. And from left to right column followed as: 2002, 2004, 2007 and 2010 outburst. The black dash line represents MJD of the first observation in the HIMS, and the black solid line represents that of the first observation in the SIMS.}
	\label{fig:5}
\end{figure*}

\begin{figure*}
	\includegraphics[width=14cm]{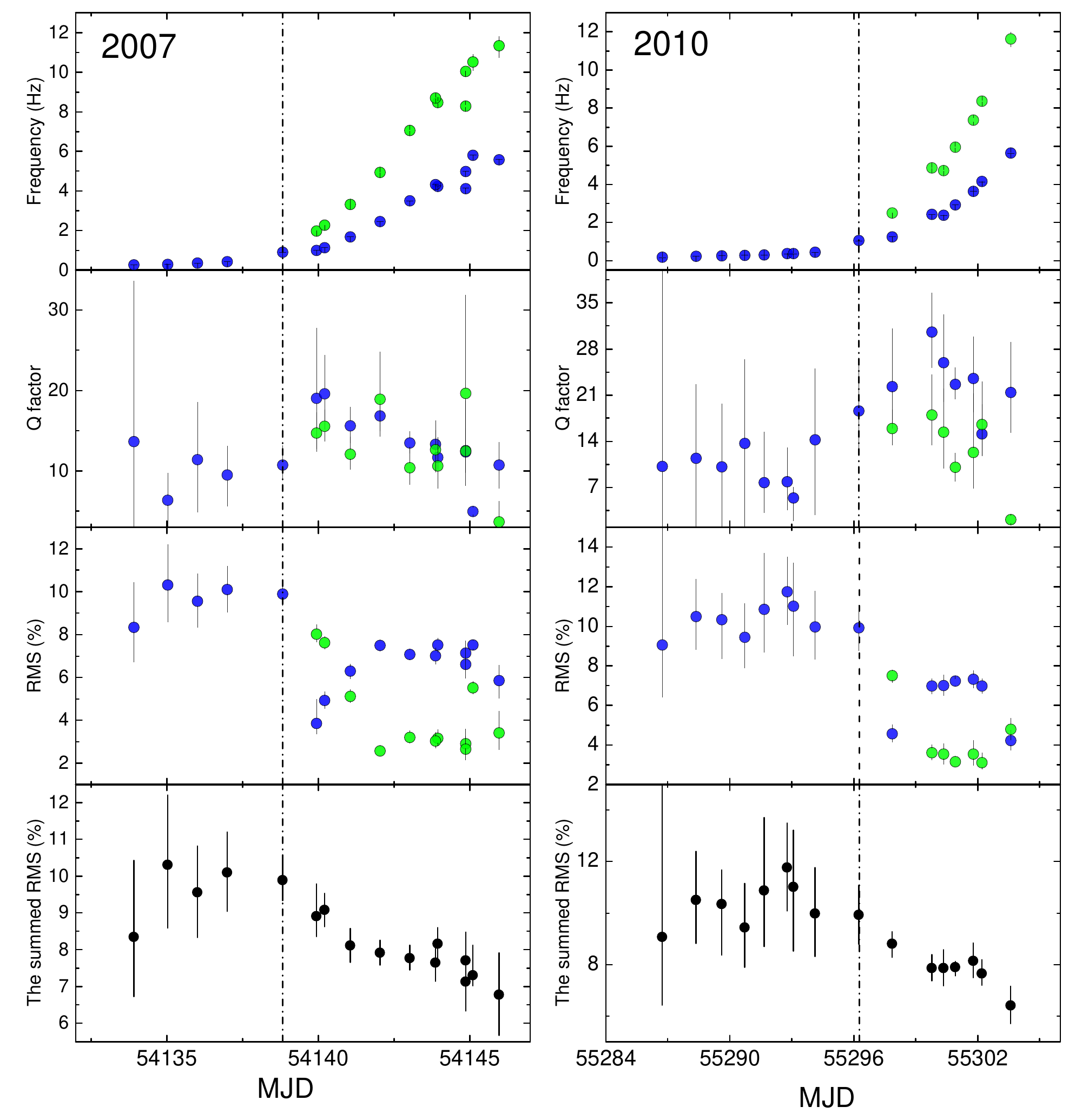}
	\caption{Evolution of type C QPO frequency, Q factor and fractional RMS (bule circles) throughout 2007 (left column) and 2010 (right column). If present, parameters of second harmonics also be plotted (green circles). And bottom panels plot the evolution of the summed QPO RMS over the fundamental and second harmonic. The dash line represents MJD of the first observation in the HIMS.}
	\label{fig:7}
\end{figure*}

\begin{figure*}
	\includegraphics[width=17cm]{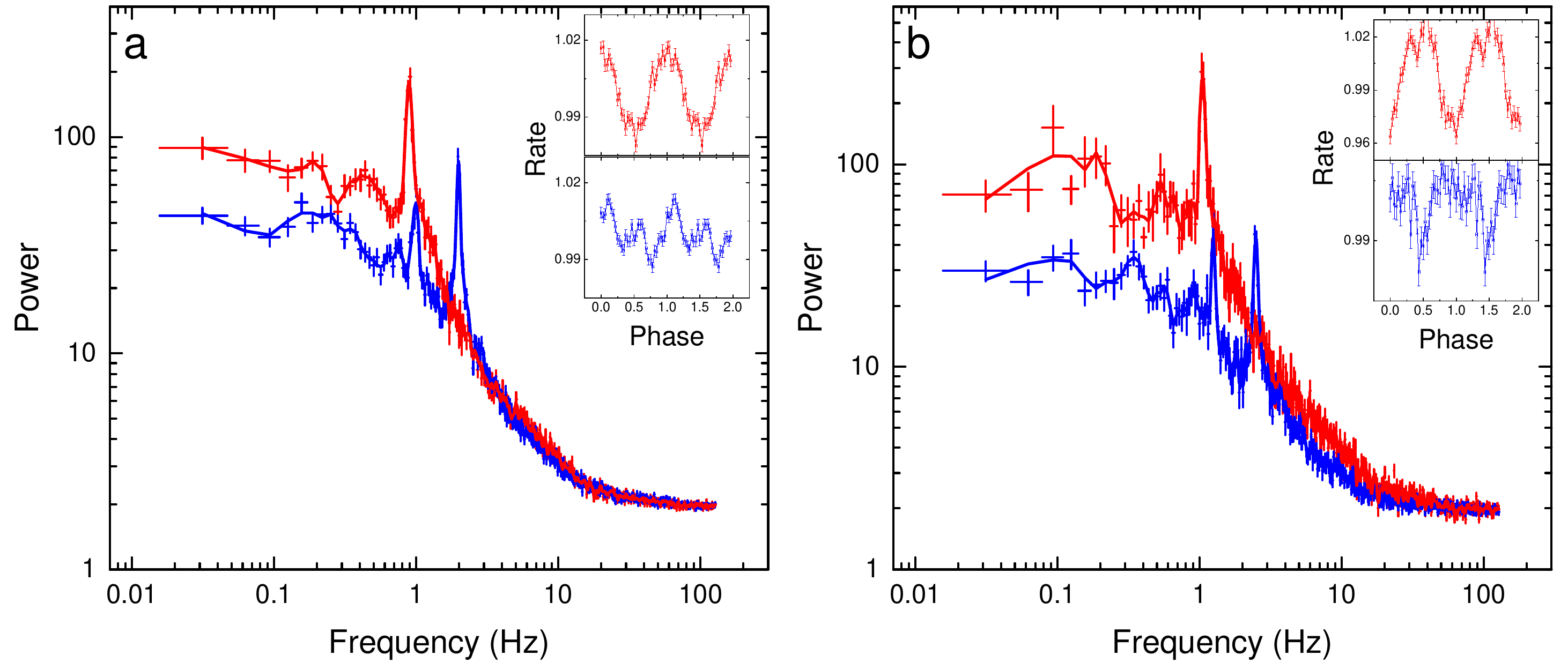}
	\caption{Examples of PDS and QPO waveform for four observations during the LHS-HIMS state transition in 2007 (a) and 2010 (b). Red: the PDS and QPO waveform of observations before the appearance of the second harmonic. Bule: the PDS and QPO waveform of the first observations which show a second harmonic.}
	\label{fig:8}
\end{figure*}

Evolutions of type C QPOs during LHS and HIMS are shown in Figure~\ref{fig:7} for the 2007 and 2010 outbursts.
The QPO frequency ($f_{\rm QPO}$) remains low with less change in LHS but increases rapidly by a magnitude  in HIMS. In the mean while, the QPO RMS evolves in an opposite manner: high in LHS and low in HIMS.  
The parameter Q shows no obvious trend of the variation in LHS but tends to decrease in HIMS. 
It is interesting that the harmonic QPO starts to show up in the HIMS and has a RMS evolution inverses to the fundamental. 
The discontinuity in fundamental RMS evolution disappears once the RMS is taken by summing over both QPOs. As shown in bottom panels of Figure~\ref{fig:7}, the summed RMS (defined as $ \sqrt{RMS_{\rm Fundamental}^2 + RMS_{\rm Harmonic}^2}$) evolves smoothly from LHS to HIMS. The comparisons between the fundamental and the harmonic are shown in Figure~\ref{fig:8} for four typical observations around the LHS-HIMS transitions in 2007 and 2010 outbursts (Obs.ID 92035-01-02-07 and 92035-01-02-06 for 2007 outburst; Obs.ID 95409-01-14-01 and 95409-01-14-02 for 2010 outburst). The QPO waveforms are folded with the centroid QPO frequencies measured in the PDS. In Figure~\ref{fig:8}, it is obvious that the harmonic in HIMS has a double-peak waveform, and an amplitude larger than that of the fundamental.


\subsection{Correlation between QPO and spectral components}
\begin{figure*}
	\includegraphics[width=17cm]{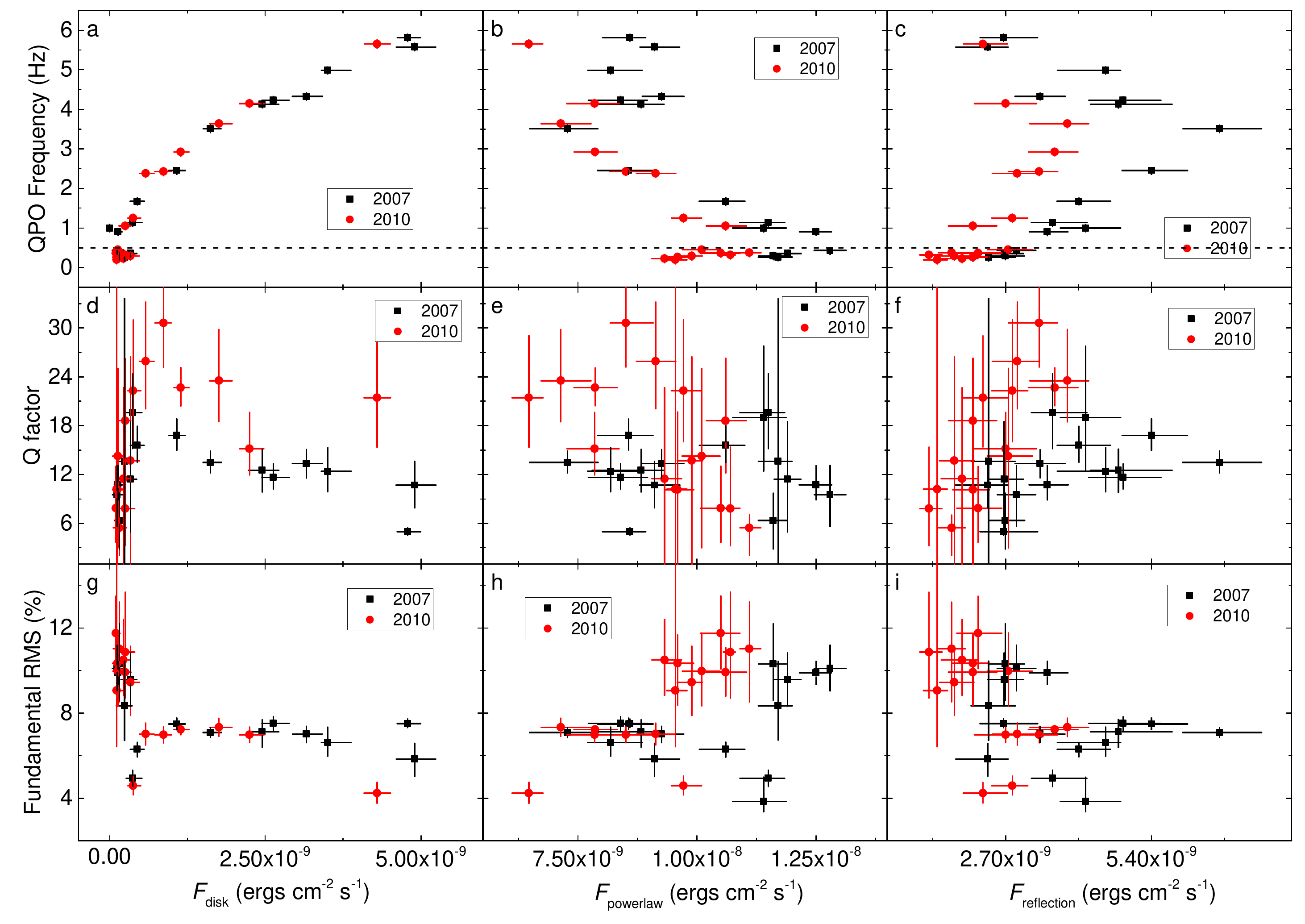}
	\caption{Type C QPO parameters (frequency, Q factor and fundamental fractional RMS) versus spectral component fluxes throughout 2007 (black squares) and 2010 (red circles) outburst. }
	\label{fig:10}
\end{figure*}

\begin{figure*}
	\includegraphics[width=16cm]{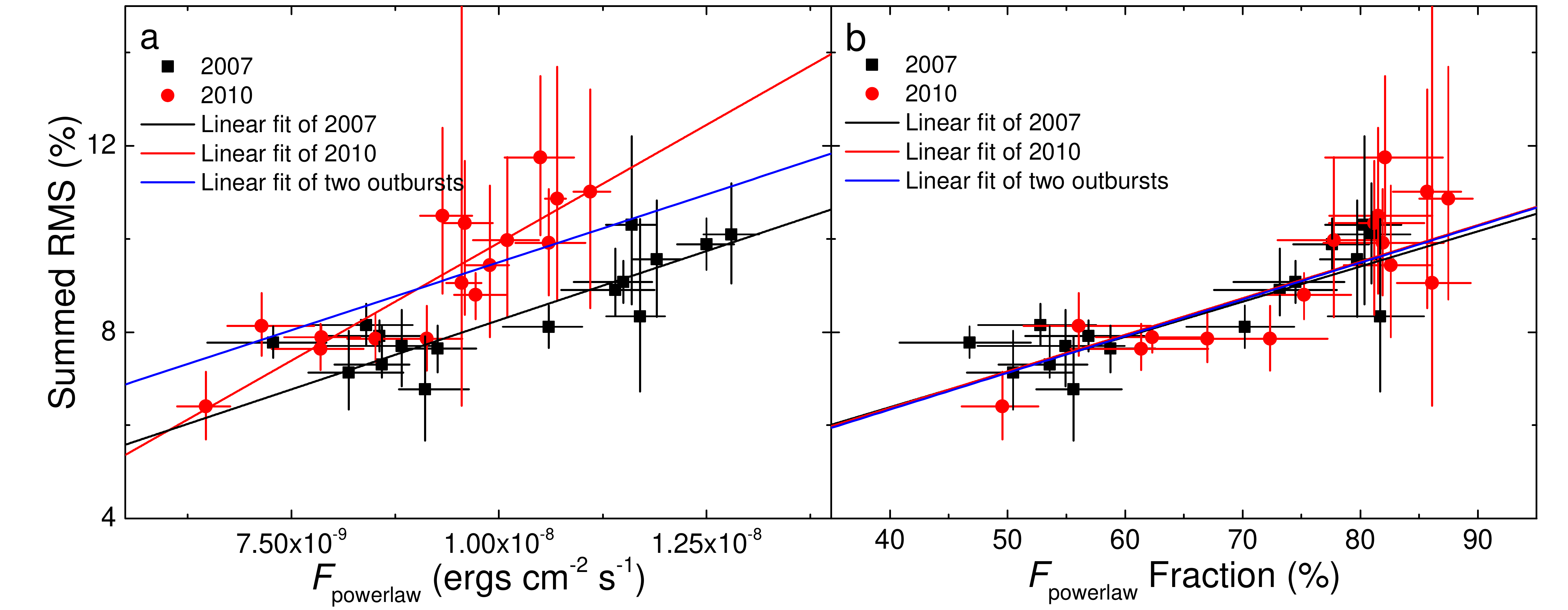}
	\caption{The summed QPO fundamental and harmonic RMS as a function of $F_{\rm pl}$ (a) and $F_{\rm pl}$ fraction (b). The linear fitting results (b) in slope/intercept of $0.076 \pm 0.015$/$3.36 \pm 1.03$, $0.079 \pm 0.030$/$3.23 \pm 2.42$ and $0.079 \pm 0.014$/$3.19 \pm 1.08$ for outburst 2007, 2010 and both, respectively.}
	\label{fig:11}
\end{figure*}

The relations between QPO parameters and different spectral components are presented in Figure~\ref{fig:10}, where the dependences of QPO frequency on $F_{\rm disk}$ and $F_{\rm pl}$ are similar to that reported in \citet{2011MNRAS.418.2292M}. The dash lines in the top three panels of Figure~\ref{fig:10} represent $f_{\rm QPO}$ = 0.5 Hz, with which the LHS and HIMS are distinguished. 
During LHS and HIMS, the QPO frequency has a positive correlation with $F_{\rm disk}$, but shows diversity with $F_{\rm pl}$ positive in LHS and negative in HIMS. 
As for the other QPO parameters, marginal correlations are visible only in Q factor against $F_{\rm disk}$, and fundamental QPO RMS against $F_{\rm pl}$ and $F_{\rm ref}$ (Figure~\ref{fig:10}d, h and i). 
The summed RMS, with the harmonic RMS set to zero when the harmonic is not detectable, shows strong correlations with $F_{\rm pl}$ in both 2007 and 2010 outbursts (Figure~\ref{fig:11}a), but with different slopes born out of a linear fitting ($(5.94 \pm 1.10)\times 10^8$, $(10.12 \pm 1.81)\times 10^8$ and $(5.63 \pm 1.47)\times 10^8$ for outburst 2007, 2010 and both).
Interestingly, if the RMS is plotted in its definition, e.g. against $F_{\rm pl}$ fraction ($ F_{\rm pl}/(F_{\rm pl}+F_{\rm disk}+F_{\rm ref})$) by supposing a non-thermal QPO origin, the RMS data from both outbursts can be well covered with a single linear function (Figure~\ref{fig:11}b).
A linear fitting results in slope/intercept of $0.076 \pm 0.015$/$3.36 \pm 1.03$, $0.079 \pm 0.030$/$3.23 \pm 2.42$ and $0.079 \pm 0.014$/$3.19 \pm 1.08$ for outburst 2007, 2010 and both, respectively.

\begin{figure}
	\includegraphics[width=\columnwidth]{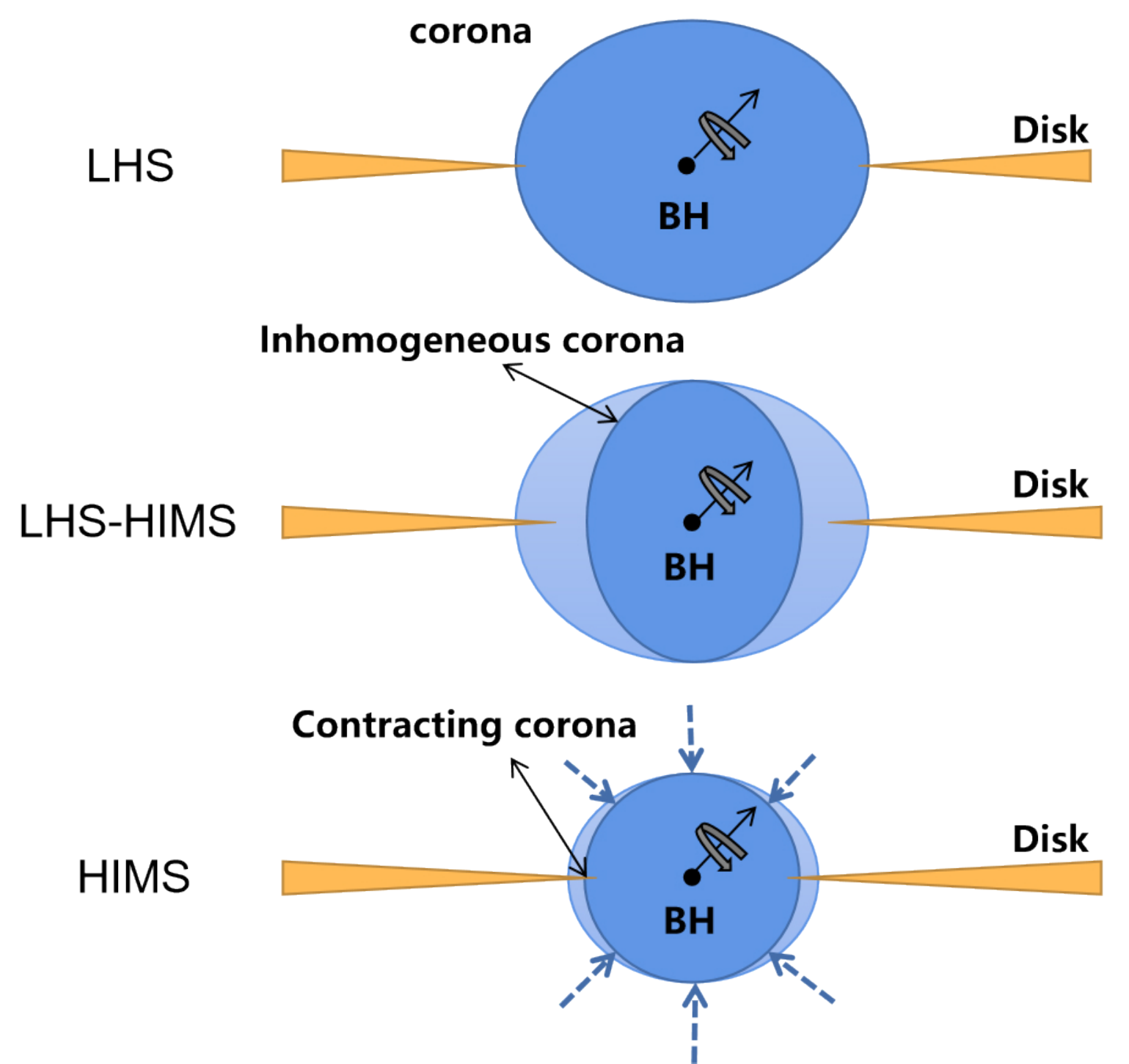}
	\caption{The simplified sketches of the evolution picture within the LHS and HIMS. LHS: the coronal scale is very large, and its structure is homogeneous. LHS-HIMS transition: disk photons firstly cool the outer part of the corona and make the corona inhomogeneous. HIMS: the accretion disk moves inward and cool the corona consistently, producing the coronal contracting.}
	\label{fig:12}
\end{figure}

\section{DISCUSSION}
\label{sec:4}
We have performed systemically analyses upon the four outbursts of GX 339-4 in spectral and timing domains, where  similarities/diversities are found among them. 
The spectral index is found to stay around 1.5 in LHS and 2.4 in SIMS/HSS. The reflection component increases all the way along with the outburst evolution until the HIMS-SIMS transition, where a turn-over shows up later than that of the hard component. The QPO RMS summed over the fundamental and harmonic manifests the same linear relation against the flux ratio of the non-thermal to the total, regardless of the individual outbursts. These findings provide more clues to our understanding the outburst and QPO.


Outburst experiencing different ‘q’ tracks in HID is a well known but less understood phenomena. It seems that the outburst with a larger accretion rate can complete the ‘q’ track with a higher peak flux. We find from outbursts of GX 339-4 that the spectral index is roughly independent of the outburst strength: it remains roughly 1.5 in LHS and 2.4 in SIMS/HSS. Since the spectral index is usually determined by the temperature and optical depth of the corona, a softer spectrum requires cooler or optically thinner corona \citep{2009MNRAS.394..207C,2012ApJ...761..109Y,2019MNRAS.487.5335L}. 
However, the corona should be optically thick if the type C QPOs are produced via LT precession \citep{2009MNRAS.397L.101I}. Hence the harder energy spectrum can be relevant to a hotter corona. The spectral evolutions in outbursts of GX 339-4 may suggest that, the corona is likely has typical temperatures in LHS and HSS that are intrinsic to the GX 339-4 system.

The reflection component is influenced by the irradiation source flux and reflection fraction decided by the disk solid angle with respect to the central irradiation source. In the outbursts of GX 339-4, the correlation between the reflection and the non-thermal is positive in LHS, SIMS and HSS but negative in HIMS. Moreover, the reflection fraction remains increasing within the LHS and HIMS.
These require a disk solid angle evolving larger in HIMS, due to either move-in disk or contracting corona \citep{2014MNRAS.442.1767P,2019Natur.565..198K}. A contracting corona, as reported previously in MAXI J1820+070 \citep{2019Natur.565..198K}, may be also needed to understand the ionization evolution as observed in the four outbursts of GX 339-4. In all the outbursts, the ionization increases until the source enters SIMS where the non-thermal emission drops abruptly, and then starts to decrease later on. 
In a contracting corona picture, becuase of the increasing solid angle and general relativistic effects, disk can enlarge its ionization by receiving more coronal hard X-ray photons when corona is contracting \citep{2013ApJ...763...48R}. Besides, the increasing thermal radiation and disk temperature can also contribute more ionization \citep{2007MNRAS.381.1697R,2014MNRAS.442.1767P}.

Q factor of the type C QPO is usually observed to increase with frequency \citep{2002ApJ...572..392B,2009MNRAS.397L.101I,2010ApJ...714.1065R}. \citet{2011MNRAS.415.2323I} explained the existence of broad-band noises in PDS via introducing the propagating fluctuations which originate from the Magneto-Rotational Instability \citep[MRI,][]{1991ApJ...376..214B} arising from the outer radius of the inner hot flow. 
They predicted that the QPO Q factor would increase under such fluctuation: less jitter in frequency for a hot flow with smaller outer radii.
However, in the case of GX 339-4, Q factor decreases in HIMS (Figure~\ref{fig:7}), probably due to presence of the harmonic in HIMS. 

To understand the QPO harmonic showing up in HIMS is a bit more challenging. 
The smooth evolution of the summed QPO RMS in LHS and HIMS suggests that a similar QPO mechanism is at work for both the fundamental and harmonic. 
For GX 339-4, the QPO RMS increases with the luminosity in LHS, while such a relation reverses in spectral states beyond. 
According to the definition of QPO RMS, a larger QPO RMS could be relevant to either the enhanced flux variability or decrease of the total flux. However, the latter is not likely since all the spectral components evolve stronger in LHS (Figure~\ref{fig:5}). Therefore, the positive correlation in LHS may simply due to the enlarged flux variability that induce QPO.  
\citet{2016MNRAS.458.1778A} found that in GX 339-4 for type C QPO the energy spectrum of the second harmonic is softer than that of the fundamental. \citet{2014MNRAS.438..657A} and \citet{2016MNRAS.458.1778A} suggested that the Comptonization region that produces QPO is inhomogeneous and, if the second harmonic comes from the outer part of the flow, it would have a softer spectrum. In HIMS, along with cooling of the corona via the disk emission, the coronal structure may change accordingly \citep{2018ApJ...858...82Y}. 
For example, disk photons firstly cool the outer part of the corona and make the corona inhomogeneous. LT precession of the inhomogeneous corona would induce an irregular QPO waveform where the harmonic is born. This is because the outer part of corona would have a different QPO phase with respect to the central part in LT precession (see Figure~\ref{fig:12}).
After the LHS-HIMS transition, the coronal region would contract and become homogeneous again. As a result, the coronal LT presscesion can produce a regular QPO waveform and the harmonic is largely suppressed.

\begin{figure*}
	\includegraphics[width=16cm]{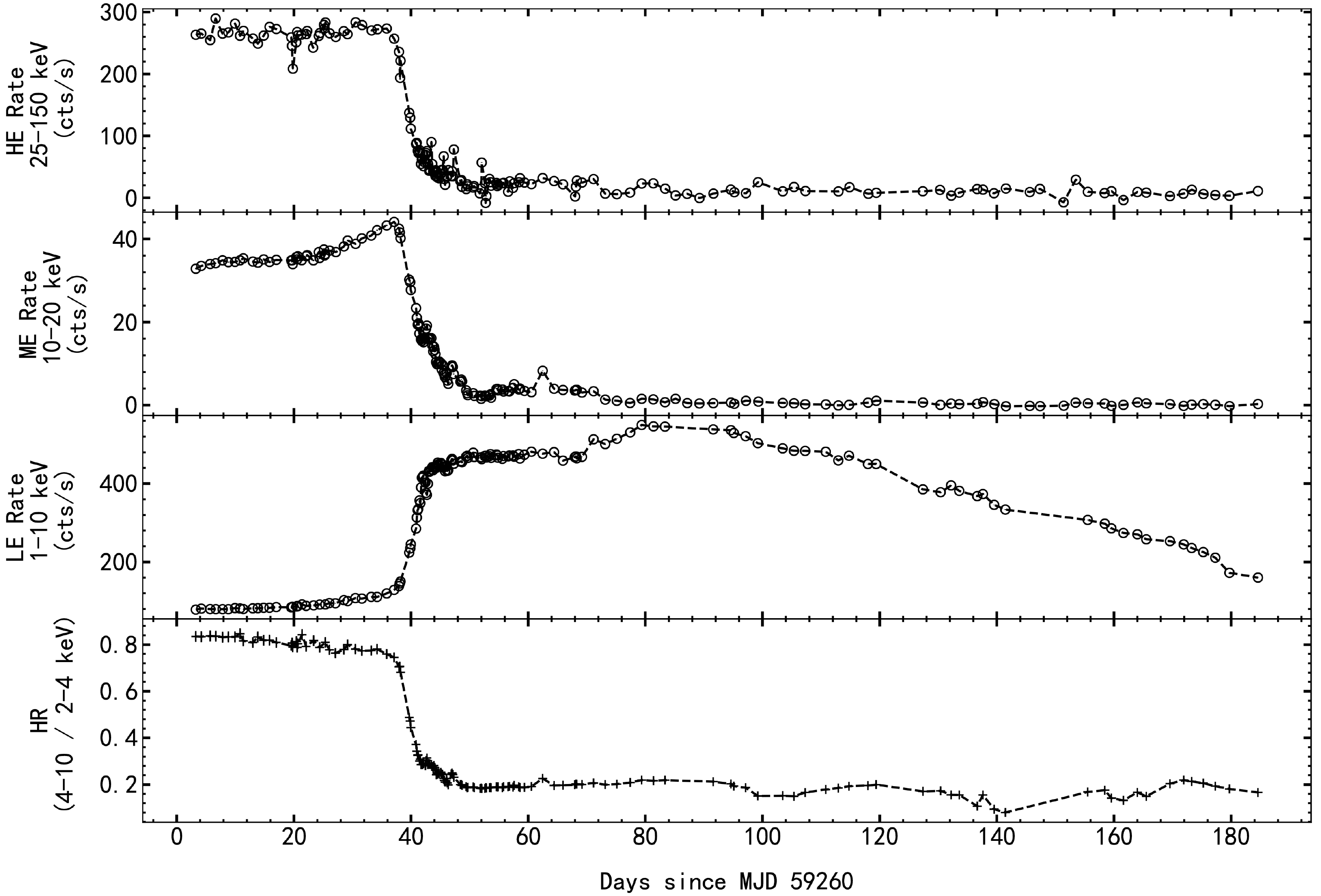}
	\caption{The 2021 outburst of GX 339-4 as observed by three telescopes (the High Energy X-ray telescope: HE, the Medium Energy X-ray telescope: ME and the Low Energy X-ray telescope: LE) of Insight-HXMT. The top three panels plot the count rate evolution of HE, ME and LE, respectively. The bottom panel plots the evolution of the hardness ratio at energies between 4-10 keV and 2-4 keV.}
	\label{fig:lc}
\end{figure*}

The joint diagnostic in spectral and timing domains by taking abundant outburst samples from GX 339-4 allows to probe the possible QPO intrinsic properties. As shown in Figure~\ref{fig:11}a, the QPO RMS evolves both lineally, but with different slopes, against the non-thermal emissions in different outbursts. However, the QPO RMS from different outbursts share the same slope once replace the non-thermal flux with the ratio of non-thermal flux to the total (see Figure~\ref{fig:11}b), i.e. the RMS is plotted in its definition of what follows:
$${\rm RMS} = \sigma \times F_{\rm c} /F_{\rm total} $$
where $F_{\rm total}$ is the time-averaged total flux, $\sigma$ denotes the intrinsic RMS, and $F_{\rm c}$ represents the flux that contributes to QPO RMS \citep{2020JHEAp..25...29K}. This is the first observational evidence that GX 339-4 has a same intrinsic QPO RMS during different outbursts. Such a result suggests that, in different outbursts the fractional variability of the non-thermal emission keeps roughly same, and hence turns out be intrinsic to the GX 339-4 system itself. Although during outburst a variety of elements, e.g. the seed photon from disk, the optical depth and geometry size of corona/jet etc., can influence the flux of the non-thermal emission, the element that can be shared in common among different outbursts is the precession angle for a given XRB system whatever precession is at work for QPO production. We also notice that, as presented in Figure~\ref{fig:11}b, the RMS data deviate slightly from the overall linear correlation for those non-thermal fraction ranging from 65\% to 75\%, corresponding to the period when the outbursts have the strong second QPO harmonic. This may be due to the slight distortion of the geometric configuration of the central corona during the state transition as illustrated in Figure~\ref{fig:12}.
Almost parallel to this work, GX 339-4 behaves a new outburst in 2021, which is monitored thoroughly by \textit{Insight}-HXMT in a broad energy band (Figure~\ref{fig:lc}), and the preliminary analyses show similar result upon QPO RMS evolution. The \textit{Insight}-HXMT observations upon the 2021 outburst will be reported elsewhere in a separate paper by the \textit{Insight}-HXMT team.

\section{CONCLUSION}
\label{sec:5}
We have analyzed four outbursts of GX 339-4 observed by \textit{RXTE}. The results born out of these analyses and comparisons suggest that, the corona may has properties (e.g. temperature) intrinsic to the GX 339-4 system itself and its geometric configuration can be influenced via disk cooling. A contracting corona can be relevant to the evolution of reflection and a inhomogeneous corona may correspond to observation of the harmonic QPO in HIMS. The same slope that links the summed RMS and the non-thermal flux ratio suggests to understand the QPO production in LT precession scenario.

\section*{ACKNOWLEDGMENTS}
This research has made use of data obtained from the High Energy Astrophysics Science Archive Research Center (HEASARC), provided by NASA's Goddard Space Flight Center. This work is supported by the National Key R\&D Program of China (2016YFA0400800, 2016YFA0400803) and the National Natural Science Foundation of China under grants U1838201, U1938101, 11733009,10903005, 11473027, 11673023, 11873035 and U1838104.

\section*{Data Availability}
The observational data used in this article are available in the HEASARC archive at \url{https://heasarc.gsfc.nasa.gov}.

\bibliographystyle{mnras}
\input{GX339_4.bbl}
\appendix
\section{}
\begin{table*}
	\renewcommand{\arraystretch}{1.19}
	\centering
	\caption{Partial results of our spectral analyses. a: $ T_{\rm in}$ is temperature at inner disk radius in units of keV; b: $E_{\rm cut}$ is the cut-off energy of CUTOFFPL in units of keV, we fix $E_{\rm cut}$ at 300 if it is less constrained; c: $ F_{\rm disk}$ is the unabsorbed flux of the multicolour disc-blackbody component in the 2.5 $\sim$ 30 keV range in units of 10$^{-9}$ ergs/cm$^{2}$/s; d: $ F_{\rm pl}$ is the unabsorbed flux of the power law with high energy exponential rolloff component in the 2.5 $\sim$ 30 keV range in units of 10$^{-9}$ ergs/cm$^{2}$/s; e: $ F_{\rm ref}$ is the unabsorbed flux of the RELXILL model (reflection) component in the 2.5 $\sim$ 30 keV range in units of 10$^{-9}$ ergs cm$^{-2}$ s$^{-1}$.}
	\setlength{\tabcolsep}{1.7mm}{
		\begin{tabular}{lcccccccccr}
			\hline
			\#&MJD&Obs.ID& $\chi^{2}_{\rm red}$&$ T_{\rm in}^{\rm a}$&$\Gamma$&$E_{\rm cut}^{\rm b}$&$\log\xi$&$F_{\rm disk}^{\rm c}$&$F_{\rm pl}^{\rm d}$&$F_{\rm ref}^{\rm e}$\\
			\hline
			1 & 52366.61 & 60705-01-55-00 & 1.15 & $1.640_{-0.221}^{+0.223}$ & $1.400_{-0.083}^{+0.035}$ & $300^{\rm f}$ & $2.99_{-0.29}^{+0.21}$ & $0.147_{-0.047}^{+0.086}$ & $2.538_{-0.078}^{+0.053}$ & $0.162_{-0.058}^{+0.062}$ \\
			2 & 52384.15 & 40031-03-02-01 & 1.12 & $--$ & $1.646_{-0.011}^{+0.009}$ & $300^{\rm f}$ & $3.14_{-0.04}^{+0.07}$ & $--$ & $13.445_{-0.140}^{+0.132}$ & $1.946_{-0.142}^{+0.147}$ \\
			3 & 52386.03 & 40031-03-02-03 & 2.39 & $--$ & $1.659_{-0.009}^{+0.008}$ & $300^{\rm f}$ & $3.16_{-0.04}^{+0.07}$ & $--$ & $13.774_{-0.105}^{+0.110}$ & $1.940_{-0.115}^{+0.110}$ \\
			4 & 52386.93 & 40031-03-02-04 & 1.32 & $--$ & $1.667_{-0.010}^{+0.010}$ & $300^{\rm f}$ & $3.13_{-0.04}^{+0.05}$ & $--$ & $13.980_{-0.131}^{+0.131}$ & $1.933_{-0.135}^{+0.134}$ \\
			5 & 52387.65 & 70109-01-04-00 & 0.62 & $1.011_{-0.151}^{+0.195}$ & $1.504_{-0.028}^{+0.027}$ & $89.2_{-9.8}^{+13.0}$ & $3.37_{-0.04}^{+0.06}$ & $0.323_{-0.104}^{+0.137}$ & $12.952_{-0.270}^{+0.264}$ & $2.616_{-0.324}^{+0.315}$ \\
			6 & 52388.07 & 40031-03-02-05 & 0.86 & $0.789_{-0.275}^{+0.459}$ & $1.516_{-0.029}^{+0.036}$ & $95.0_{-11.8}^{+25.2}$ & $3.43_{-0.06}^{+0.14}$ & $0.173_{-0.083}^{+0.130}$ & $12.624_{-0.493}^{+0.432}$ & $3.314_{-0.454}^{+0.490}$ \\
			7 & 52389.07 & 40031-03-02-06 & 0.87 & $--$ & $1.656_{-0.078}^{+0.051}$ & $300^{\rm f}$ & $3.12_{-0.18}^{+0.25}$ & $--$ & $13.482_{-1.022}^{+0.941}$ & $1.906_{-0.930}^{+3.307}$ \\
			8 & 52391.40 & 70109-01-05-00 & 0.59 & $1.138_{-0.124}^{+0.145}$ & $1.419_{-0.041}^{+0.053}$ & $61.0_{-7.4}^{+12.2}$ & $3.34_{-0.06}^{+0.06}$ & $0.661_{-0.187}^{+0.199}$ & $13.180_{-0.293}^{+0.325}$ & $2.279_{-0.390}^{+0.367}$ \\
			9 & 52400.92 & 70108-03-01-00 & 0.61 & $0.984_{-0.200}^{+0.196}$ & $1.672_{-0.030}^{+0.029}$ & $76.2_{-7.8}^{+10.7}$ & $3.53_{-0.08}^{+0.15}$ & $0.344_{-0.124}^{+0.158}$ & $12.368_{-0.283}^{+0.224}$ & $3.062_{-0.269}^{+0.355}$ \\
			10 & 52405.63 & 70109-04-01-00 & 0.52 & $0.846_{-0.020}^{+0.026}$ & $2.371_{-0.033}^{+0.039}$ & $172.8_{-42.2}^{+92.9}$ & $4.26_{-0.16}^{+0.08}$ & $4.067_{-0.205}^{+0.214}$ & $8.342_{-0.261}^{+0.327}$ & $2.422_{-0.366}^{+0.250}$ \\
			11 & 52406.08 & 70109-04-01-02 & 0.73 & $0.854_{-0.023}^{+0.027}$ & $2.291_{-0.057}^{+0.052}$ & $101.0_{-26.3}^{+50.2}$ & $4.10_{-0.20}^{+0.19}$ & $4.241_{-0.242}^{+0.239}$ & $8.168_{-0.383}^{+0.401}$ & $2.317_{-0.487}^{+0.465}$ \\
			12 & 52410.54 & 70110-01-12-00 & 0.74 & $0.855_{-0.013}^{+0.016}$ & $2.306_{-0.088}^{+0.065}$ & $91.0_{-28.9}^{+44.4}$ & $4.30_{-0.21}^{+0.11}$ & $6.666_{-0.227}^{+0.208}$ & $5.272_{-0.485}^{+0.379}$ & $1.971_{-0.393}^{+0.514}$ \\
			13 & 52411.65 & 70109-01-07-00 & 0.76 & $0.879_{-0.010}^{+0.013}$ & $2.481_{-0.073}^{+0.071}$ & $111.9_{-35.4}^{+73.0}$ & $3.99_{-0.26}^{+0.21}$ & $7.901_{-0.218}^{+0.215}$ & $4.912_{-0.176}^{+0.184}$ & $1.125_{-0.212}^{+0.222}$ \\
			14 & 52414.36 & 40031-03-03-04 & 0.86 & $0.807_{-0.008}^{+0.016}$ & $2.456_{-0.107}^{+0.037}$ & $300^{\rm f}$ & $4.04_{-0.22}^{+0.23}$ & $7.814_{-0.086}^{+0.313}$ & $2.096_{-0.203}^{+0.155}$ & $1.105_{-0.318}^{+0.213}$ \\
			15 & 52426.65 & 70110-01-17-00 & 0.81 & $0.798_{-0.014}^{+0.017}$ & $2.574_{-0.059}^{+0.073}$ & $300^{\rm f}$ & $4.06_{-0.26}^{+0.29}$ & $5.088_{-0.188}^{+0.173}$ & $2.692_{-0.214}^{+0.176}$ & $0.928_{-0.203}^{+0.253}$ \\
			16 & 52487.66 & 70109-01-17-00 & 1.07 & $0.883_{-0.007}^{+0.005}$ & $2.702_{-0.159}^{+0.083}$ & $300^{\rm f}$ & $3.32_{-0.23}^{+0.05}$ & $12.713_{-0.117}^{+0.164}$ & $1.047_{-0.032}^{+0.127}$ & $0.545_{-0.180}^{+0.065}$ \\
			17 & 52532.76 & 70110-01-47-00 & 0.89 & $0.863_{-0.010}^{+0.020}$ & $2.560_{-0.048}^{+0.035}$ & $300^{\rm f}$ & $4.31_{-0.44}^{+0.09}$ & $7.508_{-0.141}^{+0.321}$ & $4.420_{-0.220}^{+0.260}$ & $1.364_{-0.403}^{+0.203}$ \\
			18 & 52652.17 & 70109-01-31-00 & 0.92 & $0.744_{-0.011}^{+0.007}$ & $3.464_{-0.206}^{+1.992}$ & $300^{\rm f}$ & $4.17_{-0.90}^{+0.21}$ & $4.635_{-0.220}^{+0.076}$ & $--$ & $0.101_{-0.033}^{+0.063}$ \\
			19 & 52688.30 & 70109-01-36-02 & 0.59 & $0.577_{-0.009}^{+0.012}$ & $2.673_{-0.372}^{+0.502}$ & $300^{\rm f}$ & $3.44_{-0.46}^{+0.47}$ & $1.099_{-0.043}^{+0.020}$ & $0.120_{-0.033}^{+0.061}$ & $0.067_{-0.031}^{+0.023}$ \\
			20 & 52693.73 & 70110-01-86-00 & 0.49 & $0.581_{-0.016}^{+0.019}$ & $2.218_{-0.096}^{+0.183}$ & $300^{\rm f}$ & $3.80_{-0.35}^{+0.43}$ & $0.762_{-0.030}^{+0.021}$ & $0.252_{-0.060}^{+0.077}$ & $0.221_{-0.073}^{+0.066}$ \\
			21 & 52724.23 & 70110-01-94-00 & 0.83 & $0.525_{-0.005}^{+0.105}$ & $2.176_{-0.103}^{+0.026}$ & $300^{\rm f}$ & $3.83_{-0.29}^{+0.29}$ & $0.294_{-0.010}^{+0.053}$ & $0.498_{-0.111}^{+0.099}$ & $0.414_{-0.136}^{+0.083}$ \\
			22 & 52737.25 & 70110-01-98-00 & 0.70 & $0.600_{-0.257}^{+0.374}$ & $1.661_{-0.040}^{+0.066}$ & $300^{\rm f}$ & $3.27_{-0.23}^{+0.26}$ & $0.027_{-0.011}^{+0.013}$ & $0.884_{-0.059}^{+0.060}$ & $0.150_{-0.067}^{+0.062}$ \\
			\hline
			23 & 53068.32 & 80102-04-61-00 & 0.76 & $--$ & $1.470_{-0.061}^{+0.050}$ & $300^{\rm f}$ & $3.06_{-0.19}^{+0.26}$ & $--$ & $1.196_{-0.052}^{+0.061}$ & $0.186_{-0.075}^{+0.061}$ \\
			24 & 53197.11 & 90418-01-01-01 & 1.00 & $0.771_{-0.264}^{+0.257}$ & $1.476_{-0.034}^{+0.050}$ & $162.0_{-36.9}^{+86.6}$ & $3.33_{-0.14}^{+0.08}$ & $0.046_{-0.020}^{+0.018}$ & $2.124_{-0.095}^{+0.101}$ & $0.458_{-0.104}^{+0.100}$ \\
			25 & 53197.11 & 60705-01-66-00 & 1.00 & $0.771_{-0.264}^{+0.257}$ & $1.476_{-0.034}^{+0.050}$ & $162.0_{-36.9}^{+86.6}$ & $3.33_{-0.14}^{+0.08}$ & $0.046_{-0.020}^{+0.018}$ & $2.124_{-0.095}^{+0.101}$ & $0.458_{-0.104}^{+0.100}$ \\
			26 & 53215.30 & 60705-01-67-01 & 1.19 & $0.778_{-0.420}^{+0.541}$ & $1.605_{-0.048}^{+0.021}$ & $300^{\rm f}$ & $3.09_{-0.05}^{+0.17}$ & $0.041_{-0.024}^{+0.066}$ & $3.307_{-0.085}^{+0.083}$ & $0.506_{-0.127}^{+0.102}$ \\
			27 & 53218.11 & 60705-01-68-00 & 0.77 & $0.922_{-0.287}^{+0.430}$ & $1.562_{-0.059}^{+0.063}$ & $200.5_{-59.6}^{+179.3}$ & $3.17_{-0.11}^{+0.17}$ & $0.086_{-0.043}^{+0.070}$ & $3.434_{-0.115}^{+0.123}$ & $0.61_{-0.153}^{+0.130}$ \\
			28 & 53222.25 & 60705-01-68-01 & 1.61 & $--$ & $1.667_{-0.019}^{+0.020}$ & $300^{\rm f}$ & $3.46_{-0.10}^{+0.19}$ & $--$ & $3.547_{-0.210}^{+0.224}$ & $1.306_{-0.229}^{+0.219}$ \\
			29 & 53225.41 & 60705-01-69-00 & 0.89 & $0.379_{-0.287}^{+0.588}$ & $1.704_{-0.021}^{+0.017}$ & $300^{\rm f}$ & $3.40_{-0.07}^{+0.15}$ & $0.033_{-0.019}^{+0.027}$ & $3.841_{-0.232}^{+0.200}$ & $1.186_{-0.219}^{+0.243}$ \\
			30 & 53226.45 & 90704-01-01-00 & 0.62 & $0.498_{-0.193}^{+0.346}$ & $1.720_{-0.015}^{+0.010}$ & $300^{\rm f}$ & $3.39_{-0.06}^{+0.11}$ & $0.049_{-0.015}^{+0.030}$ & $3.908_{-0.180}^{+0.172}$ & $1.212_{-0.199}^{+0.190}$ \\
			31 & 53232.36 & 90110-02-01-02 & 1.10 & $0.685_{-0.030}^{+0.024}$ & $2.186_{-0.019}^{+0.024}$ & $300^{\rm f}$ & $4.06_{-0.08}^{+0.10}$ & $1.080_{-0.055}^{+0.043}$ & $1.844_{-0.255}^{+0.203}$ & $1.450_{-0.201}^{+0.280}$ \\
			32 & 53232.52 & 90110-02-01-00 & 0.59 & $0.676_{-0.016}^{+0.018}$ & $2.226_{-0.021}^{+0.019}$ & $300^{\rm f}$ & $4.10_{-0.07}^{+0.08}$ & $1.209_{-0.032}^{+0.037}$ & $1.724_{-0.155}^{+0.159}$ & $1.395_{-0.178}^{+0.168}$ \\
			33 & 53233.00 & 90110-02-01-03 & 0.87 & $0.711_{-0.016}^{+0.052}$ & $2.396_{-0.138}^{+0.036}$ & $300^{\rm f}$ & $4.04_{-0.18}^{+0.21}$ & $2.117_{-0.055}^{+0.287}$ & $1.615_{-0.309}^{+0.101}$ & $1.055_{-0.267}^{+0.206}$ \\
			34 & 53233.41 & 90704-01-02-00 & 0.60 & $0.731_{-0.014}^{+0.020}$ & $2.326_{-0.042}^{+0.028}$ & $300^{\rm f}$ & $4.31_{-0.14}^{+0.06}$ & $2.172_{-0.056}^{+0.088}$ & $1.436_{-0.174}^{+0.170}$ & $1.358_{-0.200}^{+0.164}$ \\
			35 & 53239.24 & 90704-01-05-00 & 0.44 & $0.688_{-0.013}^{+0.013}$ & $2.333_{-0.062}^{+0.061}$ & $300^{\rm f}$ & $3.78_{-0.18}^{+0.25}$ & $2.266_{-0.061}^{+0.048}$ & $0.747_{-0.086}^{+0.077}$ & $0.484_{-0.097}^{+0.115}$ \\
			36 & 53248.84 & 90704-01-09-00 & 0.72 & $0.701_{-0.008}^{+0.007}$ & $2.429_{-0.180}^{+0.288}$ & $300^{\rm f}$ & $3.42_{-0.13}^{+0.38}$ & $3.030_{-0.063}^{+0.034}$ & $0.364_{-0.052}^{+0.073}$ & $0.134_{-0.038}^{+0.043}$ \\
			37 & 53300.60 & 60705-01-80-00 & 0.88 & $0.782_{-0.006}^{+0.008}$ & $2.788_{-0.785}^{+0.179}$ & $300^{\rm f}$ & $3.31_{-1.65}^{+0.04}$ & $6.345_{-0.030}^{+0.088}$ & $0.122_{-0.007}^{+0.076}$ & $0.157_{-0.112}^{+0.020}$ \\
			38 & 53331.09 & 60705-01-84-01 & 0.52 & $0.770_{-0.008}^{+0.010}$ & $2.318_{-0.085}^{+0.108}$ & $300^{\rm f}$ & $3.70_{-0.31}^{+0.30}$ & $5.168_{-0.083}^{+0.072}$ & $1.084_{-0.125}^{+0.104}$ & $0.394_{-0.105}^{+0.142}$ \\
			39 & 53419.77 & 60705-01-89-00 & 0.44 & $0.662_{-0.005}^{+0.007}$ & $3.426_{-0.678}^{+0.608}$ & $300^{\rm f}$ & $4.02_{-1.69}^{+0.34}$ & $2.234_{-0.038}^{+0.041}$ & $0.068_{-0.033}^{+0.052}$ & $0.041_{-0.026}^{+0.016}$ \\
			40 & 53463.20 & 91105-04-09-00 & 0.52 & $0.579_{-0.041}^{+0.048}$ & $2.715_{-0.330}^{+0.417}$ & $300^{\rm f}$ & $3.49_{-0.59}^{+0.78}$ & $0.715_{-0.127}^{+0.067}$ & $0.489_{-0.118}^{+0.180}$ & $0.190_{-0.098}^{+0.084}$ \\
			41 & 53474.24 & 91105-04-12-00 & 0.51 & $0.450_{-0.121}^{+0.403}$ & $1.802_{-0.063}^{+0.028}$ & $300^{\rm f}$ & $3.36_{-0.18}^{+0.32}$ & $0.040_{-0.010}^{+0.035}$ & $1.167_{-0.094}^{+0.111}$ & $0.390_{-0.128}^{+0.096}$ \\
			42 & 53487.35 & 90704-01-13-02 & 0.63 & $1.391_{-0.255}^{+0.316}$ & $1.388_{-0.110}^{+0.081}$ & $300^{\rm f}$ & $3.06_{-0.21}^{+0.37}$ & $0.042_{-0.017}^{+0.023}$ & $0.548_{-0.029}^{+0.021}$ & $0.041_{-0.024}^{+0.032}$ \\
			\hline
			43 & 54131.13 & 92035-01-01-02 & 0.89 & $0.737_{-0.331}^{+0.306}$ & $1.549_{-0.029}^{+0.034}$ & $127.1_{-19.7}^{+34.6}$ & $3.39_{-0.05}^{+0.09}$ & $0.150_{-0.080}^{+0.095}$ & $10.730_{-0.313}^{+0.291}$ & $2.474_{-0.316}^{+0.353}$ \\
			44 & 54133.95 & 92035-01-02-01 & 0.91 & $0.844_{-0.306}^{+0.238}$ & $1.553_{-0.028}^{+0.042}$ & $109.0_{-14.5}^{+33.3}$ & $3.35_{-0.04}^{+0.08}$ & $0.241_{-0.134}^{+0.122}$ & $11.680_{-0.409}^{+0.300}$ & $2.378_{-0.325}^{+0.497}$ \\
			45 & 54135.05 & 92035-01-02-02 & 0.62 & $0.676_{-0.392}^{+0.268}$ & $1.562_{-0.019}^{+0.038}$ & $100.9_{-9.8}^{+30.0}$ & $3.39_{-0.04}^{+0.08}$ & $0.157_{-0.075}^{+0.070}$ & $11.644_{-0.312}^{+0.283}$ & $2.689_{-0.299}^{+0.358}$ \\
			46 & 54136.04 & 92035-01-02-03 & 1.15 & $0.946_{-0.225}^{+0.188}$ & $1.522_{-0.023}^{+0.048}$ & $81.3_{-7.3}^{+21.3}$ & $3.40_{-0.05}^{+0.08}$ & $0.339_{-0.155}^{+0.099}$ & $11.851_{-0.285}^{+0.287}$ & $2.672_{-0.306}^{+0.360}$ \\
			47 & 54137.02 & 92035-01-02-04 & 0.63 & $0.639_{-0.415}^{+0.374}$ & $1.616_{-0.026}^{+0.034}$ & $113.5_{-15.8}^{+31.9}$ & $3.41_{-0.05}^{+0.11}$ & $0.123_{-0.077}^{+0.104}$ & $12.842_{-0.337}^{+0.332}$ & $2.897_{-0.357}^{+0.354}$ \\
			48 & 54138.85 & 92035-01-02-07 & 0.79 & $0.519_{-0.368}^{+0.362}$ & $1.659_{-0.020}^{+0.029}$ & $97.7_{-8.8}^{+22.5}$ & $3.53_{-0.08}^{+0.16}$ & $0.133_{-0.057}^{+0.073}$ & $12.496_{-0.354}^{+0.332}$ & $3.471_{-0.335}^{+0.384}$ \\
			49 & 54139.96 & 92035-01-02-06 & 1.00 & $--$ & $1.814_{-0.014}^{+0.013}$ & $140.9_{-10.6}^{+13.8}$ & $3.85_{-0.07}^{+0.13}$ & $--$ & $11.419_{-0.651}^{+0.474}$ & $4.184_{-0.470}^{+0.649}$ \\
			50 & 54140.23 & 92035-01-03-00 & 0.74 & $0.727_{-0.153}^{+0.192}$ & $1.750_{-0.038}^{+0.032}$ & $76.3_{-10.7}^{+15.2}$ & $3.72_{-0.06}^{+0.08}$ & $0.370_{-0.108}^{+0.154}$ & $11.513_{-0.599}^{+0.346}$ & $3.570_{-0.392}^{+0.639}$ \\
			51 & 54141.08 & 92035-01-03-01 & 0.87 & $0.616_{-0.125}^{+0.128}$ & $1.888_{-0.031}^{+0.034}$ & $91.8_{-12.7}^{+22.6}$ & $3.87_{-0.07}^{+0.13}$ & $0.446_{-0.114}^{+0.111}$ & $10.611_{-0.553}^{+0.404}$ & $4.062_{-0.410}^{+0.586}$ \\
			\hline
			\multicolumn{11}{r}{continued on next page}\\
			\hline	
	\end{tabular}}
	\label{tab:3}
\end{table*}
\begin{table*}
	\renewcommand{\arraystretch}{1.19}
	\centering
	\setlength{\tabcolsep}{1.7mm}{
		\begin{tabular}{lccccccccccr}
			\hline
			\multicolumn{11}{l}{continued from previuos page}\\
			\hline
			\#&MJD&Obs.ID&$\chi^{2}_{\rm red}$&$ T_{\rm in}^{\rm a}$&$\Gamma$&$E_{\rm cut}$&$\log\xi$&$F_{\rm disk}^{\rm c}$&$F_{\rm pl}^{\rm d}$&$F_{\rm ref}^{\rm e}$\\
			\hline
			52 & 54142.06 & 92035-01-03-02 & 0.51 & $0.694_{-0.050}^{+0.065}$ & $1.968_{-0.027}^{+0.033}$ & $79.0_{-9.3}^{+17.8}$ & $4.13_{-0.05}^{+0.08}$ & $1.085_{-0.126}^{+0.134}$ & $8.563_{-0.647}^{+0.521}$ & $5.399_{-0.546}^{+0.668}$ \\
			53 & 54143.04 & 92035-01-03-03 & 0.73 & $0.680_{-0.038}^{+0.055}$ & $2.251_{-0.023}^{+0.017}$ & $300^{\rm f}$ & $4.48_{-0.04}^{+0.07}$ & $1.620_{-0.116}^{+0.173}$ & $7.279_{-0.795}^{+0.641}$ & $6.659_{-0.681}^{+0.784}$ \\
			54 & 54143.89 & 92428-01-04-00 & 0.90 & $0.809_{-0.027}^{+0.034}$ & $2.255_{-0.052}^{+0.052}$ & $100.1_{-22.2}^{+41.1}$ & $4.30_{-0.15}^{+0.06}$ & $3.160_{-0.222}^{+0.257}$ & $9.263_{-0.405}^{+0.464}$ & $3.343_{-0.469}^{+0.464}$ \\
			55 & 54143.96 & 92428-01-04-01 & 0.93 & $0.760_{-0.030}^{+0.044}$ & $2.287_{-0.036}^{+0.034}$ & $150.0_{-33.7}^{+62.4}$ & $4.41_{-0.06}^{+0.10}$ & $2.631_{-0.186}^{+0.260}$ & $8.401_{-0.680}^{+0.559}$ & $4.876_{-0.647}^{+0.701}$ \\
			56 & 54144.10 & 92428-01-04-02 & 0.61 & $0.781_{-0.039}^{+0.051}$ & $2.335_{-0.027}^{+0.019}$ & $300^{\rm f}$ & $4.44_{-0.06}^{+0.19}$ & $2.448_{-0.222}^{+0.269}$ & $8.832_{-1.028}^{+0.479}$ & $4.791_{-0.521}^{+1.003}$ \\
			57 & 54144.88 & 92428-01-04-03 & 0.98 & $0.814_{-0.017}^{+0.045}$ & $2.388_{-0.032}^{+0.016}$ & $300^{\rm f}$ & $4.69_{-0.25}^{+0.01}$ & $3.496_{-0.100}^{+0.379}$ & $8.194_{-0.485}^{+0.658}$ & $4.545_{-0.893}^{+0.279}$ \\
			58 & 54145.13 & 92035-01-03-05 & 0.87 & $0.847_{-0.015}^{+0.022}$ & $2.427_{-0.021}^{+0.017}$ & $300^{\rm f}$ & $4.35_{-0.10}^{+0.17}$ & $4.787_{-0.169}^{+0.209}$ & $8.593_{-0.573}^{+0.331}$ & $2.656_{-0.435}^{+0.629}$ \\
			59 & 54145.97 & 92428-01-04-04 & 0.97 & $0.869_{-0.021}^{+0.026}$ & $2.359_{-0.044}^{+0.060}$ & $125.4_{-32.5}^{+97.7}$ & $4.21_{-0.28}^{+0.15}$ & $4.901_{-0.299}^{+0.344}$ & $9.111_{-0.316}^{+0.528}$ & $2.369_{-0.601}^{+0.378}$ \\
			60 & 54146.05 & 92035-01-03-06 & 1.08 & $0.822_{-0.022}^{+0.032}$ & $2.425_{-0.025}^{+0.020}$ & $300^{\rm f}$ & $4.69_{-0.23}^{+0.01}$ & $4.205_{-0.186}^{+0.330}$ & $7.997_{-0.106}^{+0.512}$ & $4.205_{-0.717}^{+0.179}$ \\
			61 & 54147.03 & 92035-01-04-00 & 0.69 & $0.896_{-0.010}^{+0.016}$ & $2.535_{-0.036}^{+0.019}$ & $300^{\rm f}$ & $4.38_{-0.16}^{+0.18}$ & $8.334_{-0.209}^{+0.364}$ & $7.103_{-0.389}^{+0.285}$ & $2.093_{-0.495}^{+0.429}$ \\
			62 & 54148.15 & 92035-01-04-01 & 0.85 & $0.866_{-0.006}^{+0.010}$ & $2.533_{-0.105}^{+0.047}$ & $300^{\rm f}$ & $3.31_{-0.15}^{+0.09}$ & $10.587_{-0.105}^{+0.188}$ & $3.143_{-0.149}^{+0.091}$ & $0.472_{-0.122}^{+0.099}$ \\
			63 & 54153.71 & 92085-01-01-06 & 1.26 & $0.857_{-0.006}^{+0.009}$ & $2.778_{-0.169}^{+0.076}$ & $300^{\rm f}$ & $3.37_{-0.15}^{+0.06}$ & $11.555_{-0.099}^{+0.196}$ & $1.005_{-0.034}^{+0.103}$ & $0.672_{-0.227}^{+0.066}$ \\
			64 & 54158.81 & 92085-01-02-04 & 0.84 & $0.824_{-0.006}^{+0.007}$ & $2.509_{-0.149}^{+0.076}$ & $300^{\rm f}$ & $3.49_{-0.13}^{+0.20}$ & $9.969_{-0.065}^{+0.169}$ & $1.244_{-0.070}^{+0.102}$ & $0.625_{-0.168}^{+0.065}$ \\
			65 & 54165.55 & 92085-01-03-04 & 1.17 & $0.834_{-0.015}^{+0.012}$ & $2.570_{-0.035}^{+0.039}$ & $300^{\rm f}$ & $4.40_{-0.16}^{+0.24}$ & $6.910_{-0.268}^{+0.192}$ & $3.470_{-0.211}^{+0.181}$ & $1.468_{-0.299}^{+0.347}$ \\
			66 & 54178.20 & 92085-02-01-03 & 0.79 & $0.775_{-0.009}^{+0.012}$ & $2.511_{-0.048}^{+0.040}$ & $300^{\rm f}$ & $4.03_{-0.19}^{+0.21}$ & $4.560_{-0.093}^{+0.102}$ & $1.911_{-0.125}^{+0.093}$ & $0.812_{-0.135}^{+0.159}$ \\
			67 & 54230.71 & 92704-03-09-02 & 0.67 & $0.573_{-0.008}^{+0.026}$ & $2.183_{-0.071}^{+0.061}$ & $300^{\rm f}$ & $3.74_{-0.21}^{+0.18}$ & $0.688_{-0.013}^{+0.031}$ & $0.316_{-0.048}^{+0.072}$ & $0.357_{-0.087}^{+0.045}$ \\
			68 & 54236.60 & 92704-03-12-00 & 1.01 & $0.527_{-0.117}^{+0.193}$ & $1.936_{-0.067}^{+0.051}$ & $300^{\rm f}$ & $3.30_{-0.20}^{+0.24}$ & $0.093_{-0.015}^{+0.037}$ & $1.134_{-0.129}^{+0.083}$ & $0.359_{-0.098}^{+0.139}$ \\
			69 & 54246.87 & 92704-03-15-00 & 0.88 & $1.274_{-0.236}^{+0.264}$ & $1.482_{-0.068}^{+0.051}$ & $300^{\rm f}$ & $3.05_{-0.16}^{+0.25}$ & $0.047_{-0.019}^{+0.026}$ & $0.866_{-0.035}^{+0.024}$ & $0.059_{-0.025}^{+0.041}$ \\
			70 & 54247.92 & 92704-03-16-00 & 0.95 & $0.838_{-0.232}^{+0.507}$ & $1.522_{-0.084}^{+0.051}$ & $300^{\rm f}$ & $3.11_{-0.20}^{+0.31}$ & $0.023_{-0.010}^{+0.027}$ & $0.776_{-0.034}^{+0.037}$ & $0.092_{-0.051}^{+0.035}$ \\
			71 & 54248.84 & 92704-03-17-00 & 0.65 & $0.864_{-0.394}^{+0.526}$ & $1.504_{-0.066}^{+0.057}$ & $300^{\rm f}$ & $3.20_{-0.23}^{+0.42}$ & $0.017_{-0.011}^{+0.025}$ & $0.718_{-0.056}^{+0.034}$ & $0.107_{-0.047}^{+0.060}$ \\
			\hline
		    72 & 55249.53 & 95409-01-07-00 & 0.90 & $1.134_{-0.805}^{+0.366}$ & $1.458_{-0.050}^{+0.040}$ & $300^{\rm f}$ & $3.04_{-0.06}^{+0.15}$ & $0.058_{-0.043}^{+0.052}$ & $2.595_{-0.113}^{+0.057}$ & $0.306_{-0.086}^{+0.160}$ \\
		    73 & 55286.74 & 95409-01-12-04 & 0.85 & $0.722_{-0.233}^{+0.073}$ & $1.617_{-0.022}^{+0.037}$ & $186.1_{-31.9}^{+100.6}$ & $3.14_{-0.04}^{+0.08}$ & $0.114_{-0.066}^{+0.091}$ & $9.550_{-0.185}^{+0.239}$ & $1.428_{-0.253}^{+0.190}$ \\
		    74 & 55288.38 & 95409-01-13-03 & 1.12 & $0.750_{-0.155}^{+0.135}$ & $1.607_{-0.032}^{+0.037}$ & $186.4_{-46.0}^{+82.7}$ & $3.25_{-0.07}^{+0.07}$ & $0.223_{-0.079}^{+0.121}$ & $9.316_{-0.270}^{+0.356}$ & $1.889_{-0.388}^{+0.298}$ \\
		    75 & 55289.63 & 95409-01-13-00 & 1.03 & $0.769_{-0.220}^{+0.112}$ & $1.612_{-0.029}^{+0.033}$ & $187.4_{-37.6}^{+77.8}$ & $3.32_{-0.04}^{+0.08}$ & $0.128_{-0.076}^{+0.093}$ & $9.593_{-0.255}^{+0.336}$ & $2.092_{-0.370}^{+0.309}$ \\
		    76 & 55290.74 & 95409-01-13-04 & 0.90 & $0.984_{-0.140}^{+0.119}$ & $1.532_{-0.034}^{+0.040}$ & $102.9_{-14.9}^{+29.3}$ & $3.33_{-0.04}^{+0.06}$ & $0.334_{-0.111}^{+0.144}$ & $9.893_{-0.304}^{+0.230}$ & $1.748_{-0.278}^{+0.360}$ \\
		    77 & 55291.66 & 95409-01-13-02 & 0.90 & $1.342_{-0.327}^{+0.194}$ & $1.665_{-0.020}^{+0.017}$ & $300^{\rm f}$ & $3.09_{-0.04}^{+0.07}$ & $0.251_{-0.136}^{+0.152}$ & $10.700_{-0.147}^{+0.105}$ & $1.279_{-0.172}^{+0.237}$ \\
		    78 & 55292.80 & 95409-01-13-05 & 0.81 & $0.605_{-0.234}^{+0.063}$ & $1.636_{-0.021}^{+0.055}$ & $175.8_{-36.1}^{+132.4}$ & $3.27_{-0.13}^{+0.04}$ & $0.105_{-0.050}^{+0.063}$ & $10.528_{-0.430}^{+0.404}$ & $2.187_{-0.398}^{+0.439}$ \\
		    79 & 55293.10 & 95409-01-13-01 & 0.93 & $0.802_{-0.653}^{+0.166}$ & $1.639_{-0.035}^{+0.100}$ & $175.2_{-43.1}^{+121.0}$ & $3.30_{-0.28}^{+0.10}$ & $0.155_{-0.098}^{+0.118}$ & $11.089_{-0.203}^{+0.242}$ & $1.698_{-0.248}^{+0.255}$ \\
		    80 & 55294.13 & 95409-01-13-06 & 1.29 & $0.654_{-0.452}^{+0.064}$ & $1.600_{-0.018}^{+0.042}$ & $120.4_{-15.3}^{+44.5}$ & $3.49_{-0.06}^{+0.29}$ & $0.133_{-0.081}^{+0.096}$ & $10.075_{-0.412}^{+0.381}$ & $2.748_{-0.391}^{+0.445}$ \\
		    81 & 55296.25 & 95409-01-14-01 & 0.89 & $0.685_{-0.106}^{+0.068}$ & $1.682_{-0.029}^{+0.057}$ & $108.0_{-19.9}^{+45.3}$ & $3.36_{-0.11}^{+0.06}$ & $0.250_{-0.096}^{+0.092}$ & $10.591_{-0.401}^{+0.441}$ & $2.087_{-0.454}^{+0.441}$ \\
		    82 & 55297.88 & 95409-01-14-02 & 0.76 & $0.665_{-0.067}^{+0.050}$ & $1.764_{-0.032}^{+0.035}$ & $79.9_{-11.7}^{+16.0}$ & $3.63_{-0.08}^{+0.09}$ & $0.381_{-0.094}^{+0.124}$ & $9.725_{-0.260}^{+0.378}$ & $2.823_{-0.381}^{+0.295}$ \\
		    83 & 55299.79 & 95409-01-14-06 & 0.61 & $0.679_{-0.049}^{+0.037}$ & $1.995_{-0.028}^{+0.064}$ & $103.5_{-20.2}^{+62.0}$ & $3.98_{-0.15}^{+0.30}$ & $0.867_{-0.142}^{+0.126}$ & $8.508_{-0.336}^{+0.575}$ & $3.322_{-0.575}^{+0.343}$ \\
		    84 & 55300.34 & 95409-01-14-04 & 0.92 & $0.667_{-0.053}^{+0.044}$ & $1.999_{-0.031}^{+0.059}$ & $95.1_{-17.5}^{+35.1}$ & $4.01_{-0.08}^{+0.13}$ & $0.581_{-0.099}^{+0.137}$ & $9.134_{-0.403}^{+0.426}$ & $2.912_{-0.459}^{+0.415}$ \\
		    85 & 55300.94 & 95409-01-14-07 & 0.52 & $0.681_{-0.025}^{+0.035}$ & $2.129_{-0.021}^{+0.021}$ & $158.9_{-29.6}^{+44.6}$ & $4.11_{-0.16}^{+0.05}$ & $1.143_{-0.112}^{+0.139}$ & $7.857_{-0.445}^{+0.468}$ & $3.611_{-0.490}^{+0.435}$ \\
		    86 & 55301.79 & 95409-01-14-05 & 0.83 & $0.712_{-0.039}^{+0.061}$ & $2.253_{-0.030}^{+0.028}$ & $300^{\rm f}$ & $4.28_{-0.15}^{+0.06}$ & $1.758_{-0.149}^{+0.208}$ & $7.144_{-0.411}^{+0.633}$ & $3.840_{-0.696}^{+0.393}$ \\
		    87 & 55302.20 & 95409-01-15-00 & 1.16 & $0.743_{-0.032}^{+0.048}$ & $2.330_{-0.033}^{+0.025}$ & $300^{\rm f}$ & $4.13_{-0.20}^{+0.19}$ & $2.245_{-0.164}^{+0.234}$ & $7.848_{-0.588}^{+0.517}$ & $2.696_{-0.581}^{+0.570}$ \\
		    88 & 55303.61 & 95409-01-15-01 & 0.87 & $0.820_{-0.015}^{+0.017}$ & $2.323_{-0.033}^{+0.029}$ & $147.9_{-33.8}^{+50.7}$ & $4.15_{-0.08}^{+0.21}$ & $4.303_{-0.209}^{+0.215}$ & $6.474_{-0.349}^{+0.293}$ & $2.282_{-0.368}^{+0.454}$ \\
		    89 & 55304.72 & 95409-01-15-02 & 0.74 & $0.885_{-0.009}^{+0.016}$ & $2.512_{-0.038}^{+0.030}$ & $300^{\rm f}$ & $4.18_{-0.30}^{+0.21}$ & $6.969_{-0.139}^{+0.237}$ & $5.359_{-0.265}^{+0.233}$ & $1.294_{-0.325}^{+0.285}$ \\
		    90 & 55308.99 & 95409-01-15-06 & 0.87 & $0.833_{-0.011}^{+0.020}$ & $2.557_{-0.050}^{+0.026}$ & $300^{\rm f}$ & $4.41_{-0.24}^{+0.24}$ & $6.799_{-0.195}^{+0.318}$ & $4.355_{-0.386}^{+0.265}$ & $1.853_{-0.473}^{+0.390}$ \\
		    91 & 55315.71 & 95409-01-16-05 & 1.10 & $0.851_{-0.014}^{+0.015}$ & $2.456_{-0.035}^{+0.028}$ & $300^{\rm f}$ & $4.03_{-0.15}^{+0.23}$ & $5.943_{-0.172}^{+0.171}$ & $4.834_{-0.223}^{+0.117}$ & $1.240_{-0.201}^{+0.312}$ \\
		    92 & 55316.12 & 95409-01-17-00 & 0.95 & $0.832_{-0.014}^{+0.022}$ & $2.458_{-0.043}^{+0.035}$ & $300^{\rm f}$ & $4.25_{-0.33}^{+0.16}$ & $5.860_{-0.179}^{+0.254}$ & $4.598_{-0.272}^{+0.301}$ & $1.686_{-0.485}^{+0.353}$ \\
		    93 & 55321.73 & 95409-01-17-05 & 0.77 & $0.839_{-0.012}^{+0.014}$ & $2.496_{-0.039}^{+0.032}$ & $300^{\rm f}$ & $4.01_{-0.18}^{+0.27}$ & $5.654_{-0.146}^{+0.137}$ & $3.869_{-0.216}^{+0.098}$ & $0.990_{-0.152}^{+0.275}$ \\
		    94 & 55323.22 & 95409-01-18-00 & 0.74 & $0.821_{-0.013}^{+0.022}$ & $2.538_{-0.048}^{+0.028}$ & $300^{\rm f}$ & $4.36_{-0.27}^{+0.23}$ & $4.993_{-0.163}^{+0.254}$ & $4.180_{-0.315}^{+0.246}$ & $1.435_{-0.416}^{+0.359}$ \\
		    95 & 55324.20 & 95335-01-01-07 & 0.96 & $0.821_{-0.012}^{+0.020}$ & $2.485_{-0.055}^{+0.034}$ & $300^{\rm f}$ & $4.61_{-0.26}^{+0.07}$ & $4.989_{-0.136}^{+0.239}$ & $3.265_{-0.207}^{+0.376}$ & $2.098_{-0.514}^{+0.169}$ \\
		    96 & 55324.41 & 95335-01-01-01 & 0.86 & $0.809_{-0.011}^{+0.017}$ & $2.538_{-0.040}^{+0.027}$ & $300^{\rm f}$ & $4.06_{-0.20}^{+0.23}$ & $5.116_{-0.109}^{+0.173}$ & $3.642_{-0.177}^{+0.133}$ & $1.137_{-0.203}^{+0.207}$ \\
		    97 & 55326.19 & 95335-01-01-05 & 0.95 & $0.806_{-0.017}^{+0.015}$ & $2.547_{-0.045}^{+0.055}$ & $300^{\rm f}$ & $4.04_{-0.26}^{+0.27}$ & $5.162_{-0.215}^{+0.165}$ & $3.305_{-0.208}^{+0.177}$ & $0.987_{-0.209}^{+0.287}$ \\
		    98 & 55326.3 & 95335-01-01-06 & 1.17 & $0.804_{-0.011}^{+0.014}$ & $2.509_{-0.042}^{+0.033}$ & $300^{\rm f}$ & $4.03_{-0.22}^{+0.23}$ & $5.312_{-0.112}^{+0.142}$ & $2.912_{-0.156}^{+0.110}$ & $0.933_{-0.174}^{+0.215}$ \\
		    99 & 55327.27 & 95409-01-18-05 & 0.71 & $0.802_{-0.018}^{+0.021}$ & $2.592_{-0.051}^{+0.086}$ & $300^{\rm f}$ & $4.07_{-0.37}^{+0.32}$ & $4.821_{-0.246}^{+0.193}$ & $3.559_{-0.188}^{+0.238}$ & $0.841_{-0.251}^{+0.258}$ \\
		    100 & 55330.30 & 95409-01-19-00 & 1.1 & $0.788_{-0.012}^{+0.021}$ & $2.538_{-0.072}^{+0.037}$ & $300^{\rm f}$ & $4.33_{-0.30}^{+0.25}$ & $4.685_{-0.127}^{+0.223}$ & $2.738_{-0.348}^{+0.197}$ & $1.128_{-0.310}^{+0.370}$ \\
		    101 & 55603.99 & 96409-01-07-00 & 0.73 & $0.261_{-0.138}^{+1.407}$ & $1.739_{-0.086}^{+0.071}$ & $300^{\rm f}$ & $3.34_{-0.29}^{+0.50}$ & $0.013_{-0.013}^{+0.027}$ & $0.736_{-0.080}^{+0.085}$ & $0.204_{-0.098}^{+0.079}$ \\
		    102 & 55606.9 & 96409-01-07-01 & 0.94 & $1.027_{-0.238}^{+0.426}$ & $1.564_{-0.084}^{+0.076}$ & $300^{\rm f}$ & $3.18_{-0.32}^{+0.50}$ & $0.032_{-0.016}^{+0.028}$ & $0.703_{-0.048}^{+0.040}$ & $0.085_{-0.046}^{+0.046}$ \\
		    103 & 55607.77 & 96409-01-07-02 & 0.67 & $0.883_{-0.310}^{+0.426}$ & $1.566_{-0.100}^{+0.062}$ & $300^{\rm f}$ & $3.02_{-0.24}^{+0.23}$ & $0.026_{-0.014}^{+0.019}$ & $0.636_{-0.035}^{+0.028}$ & $0.073_{-0.035}^{+0.040}$ \\
			\hline
	\end{tabular}}
\end{table*}
\begin{table*}
	\renewcommand\arraystretch{1.36}
	\centering
	\caption{Timing results of outburst 2007 and 2010. Q factor is defined as $f_{\rm QPO}/\rm FWHM$ and fundamental RMS has been corrected ($ RMS=\sqrt{P}\times \left[(S+B)/S\right]$). Each column of type C QPO parameters contains the fundamental and second harmonic (e.g. $f_{\rm Fundamental}/f_{\rm Harmonic}$). We followed \citet{2011MNRAS.418.2292M} to classify the spectral state. And the number of these observations in the first column followed this in Table~\ref{tab:3}.}
	\setlength{\tabcolsep}{3.2mm}{
		\begin{tabular}{ccccccccc}
			\hline
			\#     &     Frequency(Hz)   &  Q  &   Fundamental RMS(\%)   &     QPO type   &   State     \\
			\hline
			44 &  $0.263_{-0.005}^{+0.003}/--$ &$13.63_{-20.06}^{+19.98}/--$ & $8.34_{-1.72}^{+2.19}/--$& C   & LHS   \\
			45 &  $0.291_{-0.007}^{+0.006}/--$ &$6.35_{-3.40}^{+3.36}/--$    & $10.31_{-1.83}^{+2.02}/--$& C & LHS   \\
			46 &  $0.356_{-0.004}^{+0.004}/--$ &$11.43_{-7.10}^{+7.09}/--$   & $9.56_{-1.34}^{+1.37}/--$ & C &  LHS   \\
			47 &  $0.429_{-0.006}^{+0.006}/--$ &$9.51_{-3.59}^{+3.59}/--$    & $10.10_{-1.16}^{+1.21}/--$&  C & LHS  \\
			48 &  $0.901_{-0.005}^{+0.006}/--$ &$10.74_{-2.37}^{+2.38}/--$   & $9.89_{-0.66}^{+0.67}/--$ &  C & HIMS \\
			49 &  $0.994_{-0.007}^{+0.006}/1.988_{-0.001}^{+0.001}$ &$18.78_{-7.07}^{+7.07}/14.73_{-1.96}^{+2.52}$ & $3.93_{-0.24}^{+0.44}/8.02_{-0.37}^{+0.44}$ & C &  HIMS \\   
			50 &  $1.134_{-0.005}^{+0.005}/2.269_{-0.007}^{+0.007}$ &$18.90_{-4.31}^{+4.30}/15.54_{-1.84}^{+2.05}$ & $4.98_{-0.39}^{+0.39}/7.62_{-0.29}^{+0.29}$ & C & HIMS  \\
			51 &  $1.673_{-0.007}^{+0.007}/3.334_{-0.016}^{+0.015}$ &$15.50_{-2.28}^{+2.28}/12.05_{-1.88}^{+2.20}$ & $6.27_{-0.38}^{+0.38}/5.13_{-0.29}^{+0.31}$ & C & HIMS  \\
			52 &  $2.448_{-0.006}^{+0.006}/4.940_{-0.026}^{+0.025}$ &$18.00_{-1.91}^{+1.91}/18.90_{-4.60}^{+5.87}$ & $7.32_{-0.32}^{+0.32}/2.58_{-0.23}^{+0.25}$ & C & HIMS  \\
			53 &  $3.503_{-0.010}^{+0.009}/7.064_{-0.043}^{+0.042}$ &$13.30_{-1.30}^{+1.30}/10.38_{-2.08}^{+2.59}$ & $7.16_{-0.29}^{+0.29}/3.20_{-0.27}^{+0.32}$ & C & HIMS  \\
			54 &  $4.241_{-0.017}^{+0.016}/8.706_{-0.043}^{+0.042}$ &$14.02_{-1.84}^{+1.83}/12.64_{-2.82}^{+3.57}$ & $6.95_{-0.37}^{+0.37}/3.04_{-0.33}^{+0.40}$ & C & HIMS \\
			55 &  $4.236_{-0.016}^{+0.015}/8.474_{-0.071}^{+0.074}$ &$11.62_{-1.66}^{+1.66}/10.62_{-2.83}^{+3.52}$ & $7.52_{-0.44}^{+0.43}/3.17_{-0.37}^{+0.40}$ &C  &HIMS&  \\
			56 &  $4.143_{-0.018}^{+0.017}/8.296_{-0.069}^{+0.071}$ &$12.82_{-1.86}^{+1.86}/12.48_{-4.27}^{+6.25}$ & $7.04_{-0.42}^{+0.43}/2.91_{-0.47}^{+0.66}$ & C  &  HIMS  \\
			57 &  $4.995_{-0.022}^{+0.022}/10.047_{-0.001}^{+0.001}$ &$10.98_{-2.17}^{+2.17}/19.63_{-8.18}^{+12.25}$& $7.03_{-0.42}^{+0.43}/2.66_{-0.50}^{+0.60}$ & C  &  HIMS  \\
			58 &  $5.814_{-0.023}^{+0.023}/10.510_{-0.428}^{+0.385}$ &$4.98_{-0.48}^{+0.48}/1.53_{-0.24}^{+0.22}$  & $7.51_{-0.29}^{+0.28}/5.53_{-0.28}^{+0.29}$ &C  & HIMS  \\
			59 &  $5.591_{-0.031}^{+0.030}/11.331_{-0.588}^{+0.462}$ &$9.99_{-1.76}^{+1.76}/3.69_{-1.92}^{+2.55}$  & $6.16_{-0.46}^{+0.47}/3.42_{-0.79}^{+1.01}$ &C  & HIMS& \\
			\hline
			73 &  $0.195_{-0.022}^{+0.010}/--$ &$10.20_{-61.04}^{+60.43}/--$ & $9.06_{-2.86}^{+10.92}/--$  &C  & LHS    \\
			74 &  $0.229_{-0.005}^{+0.005}/--$ &$11.47_{-11.21}^{+11.20}/--$ & $10.50_{-1.86}^{+2.08}/--$ &C  &  LHS    \\
			75 &  $0.265_{-0.004}^{+0.006}/--$ &$10.15_{-9.51}^{+9.55}/--$   & $10.34_{-2.15}^{+1.51}/--$& C  &  LHS    \\
			76 &  $0.291_{-0.005}^{+0.005}/--$ &$13.72_{-12.72}^{+12.72}/--$ & $9.44_{-1.68}^{+1.84}/--$ & C  &  LHS   \\
			77 &  $0.315_{-0.007}^{+0.006}/--$ &$7.79_{-7.63}^{+7.61}/--$    & $10.87_{-2.37}^{+3.03}/--$& C  &  LHS   \\
			78 &  $0.366_{-0.008}^{+0.010}/--$ &$7.87_{-5.09}^{+5.13}/--$    & $11.75_{-2.00}^{+2.07}/--$& C  &  LHS   \\
			79 &  $0.373_{-0.013}^{+0.017}/--$ &$5.42_{-1.56}^{+1.63}/--$    & $11.01_{-2.70}^{-2.40}/--$& C  &  LHS   \\
			80 &  $0.447_{-0.008}^{+0.005}/--$ &$14.25_{-10.85}^{+10.78}/--$ & $9.97_{-1.91}^{+2.05}/--$ & C  &  LHS    \\
			82 &  $1.053_{-0.010}^{+0.011}/--$ &$18.59_{-7.67}^{+7.68}/--$   & $9.92_{-1.43}^{+1.46}/--$ & C  &  HIMS    \\
			83 &  $1.250_{-0.007}^{+0.007}/2.494_{-0.009}^{+0.009}$ &$22.01_{-8.49}^{+8.49}/15.94_{-2.42}^{+2.88}$ & $4.58_{-0.53}^{+0.53}/7.52_{-0.36}^{+0.27}$ & C  &  HIMS   \\
			84 &  $2.428_{-0.006}^{+0.006}/4.861_{-0.031}^{+0.031}$ &$31.07_{-5.94}^{+5.94}/17.99_{-4.58}^{+6.20}$ & $6.95_{-0.49}^{+0.49}/3.62_{-0.37}^{+0.42}$ & C  &  HIMS  \\
			85 &  $2.374_{-0.010}^{+0.010}/4.730_{-0.048}^{+0.054}$ &$26.68_{-7.91}^{+7.91}/15.38_{-5.42}^{+6.59}$ & $6.97_{-0.67}^{+0.64}/3.54_{-0.49}^{+0.53}$ & C  &  HIMS   \\
			86 &  $2.921_{-0.007}^{+0.005}/5.952_{-0.041}^{+0.039}$ &$20.75_{-2.38}^{+2.37}/10.11_{-2.11}^{+2.16}$ & $7.50_{-0.37}^{+0.35}/3.17_{-0.27}^{+0.27}$ & C  &  HIMS   \\
			87 &  $3.642_{-0.011}^{+0.012}/7.372_{-0.088}^{+0.119}$ &$23.57_{-6.30}^{+6.30}/12.31_{-5.35}^{+8.99}$ & $7.32_{-0.61}^{+0.62}/3.55_{-0.58}^{+0.68}$ & C  &  HIMS  \\
			88 &  $4.146_{-0.017}^{+0.018}/8.37_{-0.054}^{+0.053}$ &$15.16_{-4.48}^{+4.48}/16.58_{-4.69}^{+6.52}$   & $6.98_{-0.57}^{+0.57}/3.11_{-0.31}^{+0.49}$ & C  &  HIMS   \\
			89 &  $5.654_{-0.021}^{+0.023}/11.628_{-0.421}^{+0.364}$ &$21.43_{-7.64}^{+7.64}/2.19_{-0.56}^{+0.60}$   & $4.24_{-0.56}^{+0.59}/4.81_{-0.53}^{+0.54}$ & C  &  HIMS   \\
			\hline
	\end{tabular}}
	\label{tab:2}
\end{table*}
\end{document}